\documentclass[a4paper,fleqn,preprint]{cas-sc}

\usepackage[numbers]{natbib}
\usepackage{graphicx}
\usepackage{dcolumn}
\usepackage{bm}
\usepackage{wrapfig}
\usepackage{colonequals}

\def\tsc#1{\csdef{#1}{\textsc{\lowercase{#1}}\xspace}}
\tsc{WGM}
\tsc{QE}
\tsc{EP}
\tsc{PMS}
\tsc{BEC}
\tsc{DE}

\begin{document}
\let\WriteBookmarks\relax
\def\floatpagepagefraction{1}
\def\textpagefraction{.001}
\shorttitle{RQA border effect corrections}
\shortauthors{Kraemer and Marwan}

\title [mode = title]{Border effect corrections for diagonal line based recurrence quantification analysis measures}

\author[1,2]{K.Hauke Kraemer}[orcid=0000-0002-9943-5391, twitter = KH_Kraemer]
\cormark[1]
\fnmark[1]
\ead{hkraemer@pik-potsdam.de}
\ead{hkraemer@uni-potsdam.de}

\credit{Conceptualization of this study, Methodology, Software, Data curation, Writing}

\address[1]{Potsdam Institute for Climate Impact Research, Member of the Leibniz Association,
Telegrafenberg A31, 14473 Potsdam, 
Germany, EU}
\address[2]{Institute of Geosciences, University of Potsdam, Karl-Liebknecht-Str. 
24-25, 14476 Potsdam-Golm, Germany, EU}

\author[1]{Norbert Marwan}
\ead{marwan@pik-potsdam.de}
\credit{Conceptualization of this study, Methodology, Software, Data curation, Writing}

\cortext[cor1]{Principal Corresponding author}

\begin{abstract}

Recurrence Quantification Analysis (RQA) defines a number of quantifiers, 
which base upon diagonal line structures in the recurrence plot (RP). Due to the
finite size of an RP, these lines can be cut by the borders of the RP and, thus, bias 
the length distribution of diagonal lines and, consequently, the line based RQA measures.
In this letter we investigate the impact of the mentioned border effects and of the thickening of diagonal
lines in an RP (caused by tangential motion) on the estimation of the diagonal line length 
distribution, quantified by its entropy. 
Although a relation to the Lyapunov spectrum is theoretically
expected, the mentioned entropy yields contradictory results in many studies. 
Here we summarize correction schemes for both, the
border effects and the tangential motion and systematically compare them to methods from the literature. We show that these corrections lead to the expected behavior of the diagonal line
length entropy, in particular meaning zero values in case of a regular motion and positive values for
chaotic motion. Moreover, we test these methods
under noisy conditions, in order to supply practical tools for applied statistical research. 
\end{abstract}

\begin{highlights}
\item In recurrence quantification analysis, border effects and tangential motion can heavily bias diagonal line based characteristics.
\item Border effect bias can be minimized by a simple alteration of the diagonal line length histogram. 
\item A parameter free, skeletonization method reduces artifacts due to tangential motion and leaves a RP of thin diagonal lines.
\item The proposed correction schemes lead to diagonal line entropy values that fall within analytically derived expectation ranges, even in the presence of noise.
\end{highlights}


\maketitle

\section{Introduction}
Recurrence quantification analysis (RQA) is a powerful tool for the identification of characteristic dynamics and  
regime changes \cite{marwan2007,webber2015}. This approach is successfully applied in many scientific disciplines 
\cite{marwan2002herz,guimaraes2008,facchini2009b,webber2009,guhathakurta2010,hirata2011b,subramaniyam2014,santos2015,kopacek2010,mitra2014,nair2014}. 
Several measures of complexity are defined on geometric features (such as diagonal and vertical lines) in the recurrence plot (RP), 
which represents time points $j$ when a state $\vec{x}_i$ at time $i$
recurs \cite{zbilut92, marwan2002herz,marwan2007,webber2015}.
These line structures represent typical dynamical behavior and are related to certain properties of the dynamical system,
e.g., chaotic or periodic dynamics. Therefore, their quantitative study by the RQA measures within sliding windows
is a frequently used task for the detection of regime changes \cite{trulla96,donges2011c,eroglu2014,webber2015}.
However, as some RQA measures rely on the probability distribution of the lengths of the diagonal lines in an RP, the artificial alteration
of these lines due to border effects \cite{censi2004,eroglu2014}, insufficient embedding \cite{marwan2007,marwan2011}, or a certain sampling
setting \cite{theiler1986,facchini2007a} can have significant impact on these measures. A few ideas have been suggested to overcome
such problems \cite{censi2004,schultz2011,wendi2018b}. Here we review these ideas, propose novel correction schemes,
and systematically compare them.

\section{Recurrence quantification analysis and border effects}\label{sec_rqa_border}

\begin{figure}
\begin{center}
\includegraphics[width=0.7\textwidth]{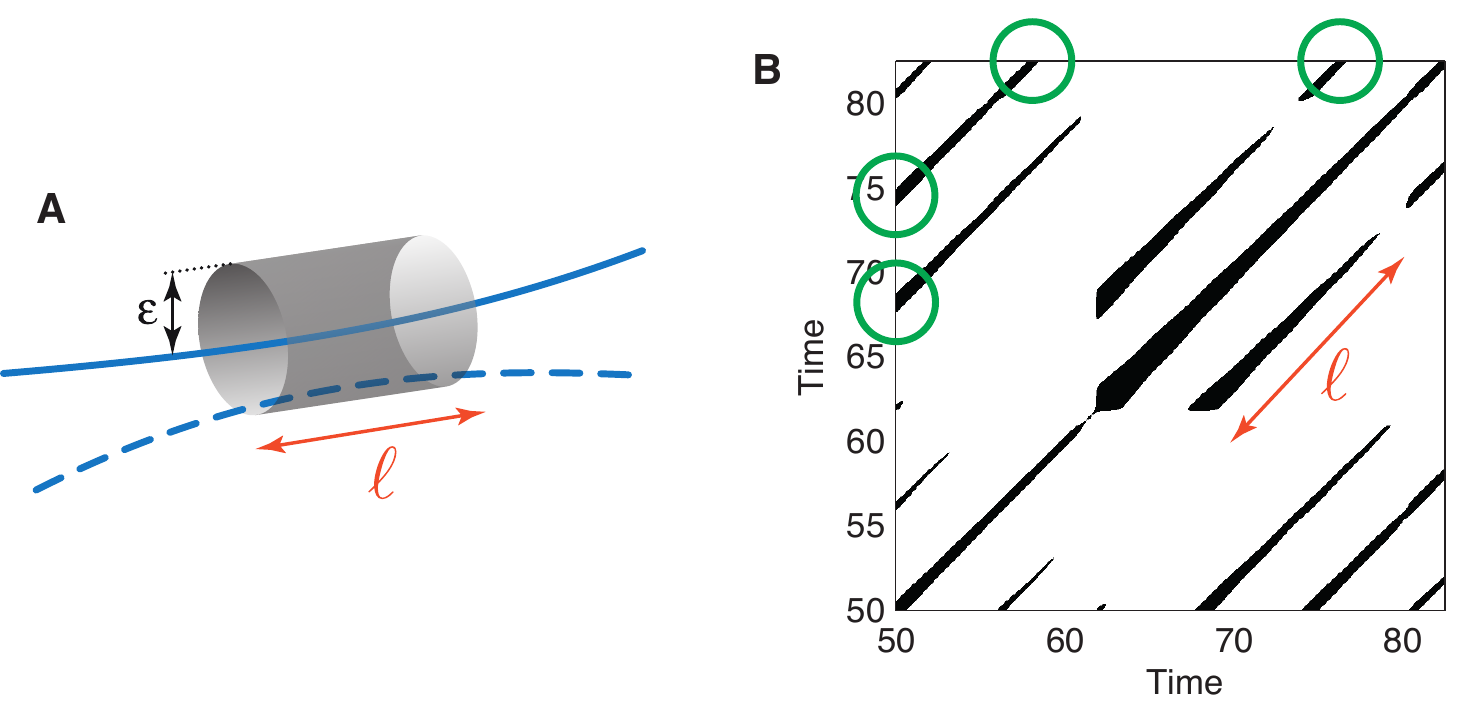}
\caption{Parallel and close parts of a phase space trajectory (A)
correspond to diagonal lines of length $\ell$ in an RP (B). Diagonal lines
can be cut by the border of the RP (green circles).}\label{fig_eps-tube}
\end{center}
\end{figure}

\begin{figure}
\centerline{\includegraphics[width=\textwidth]{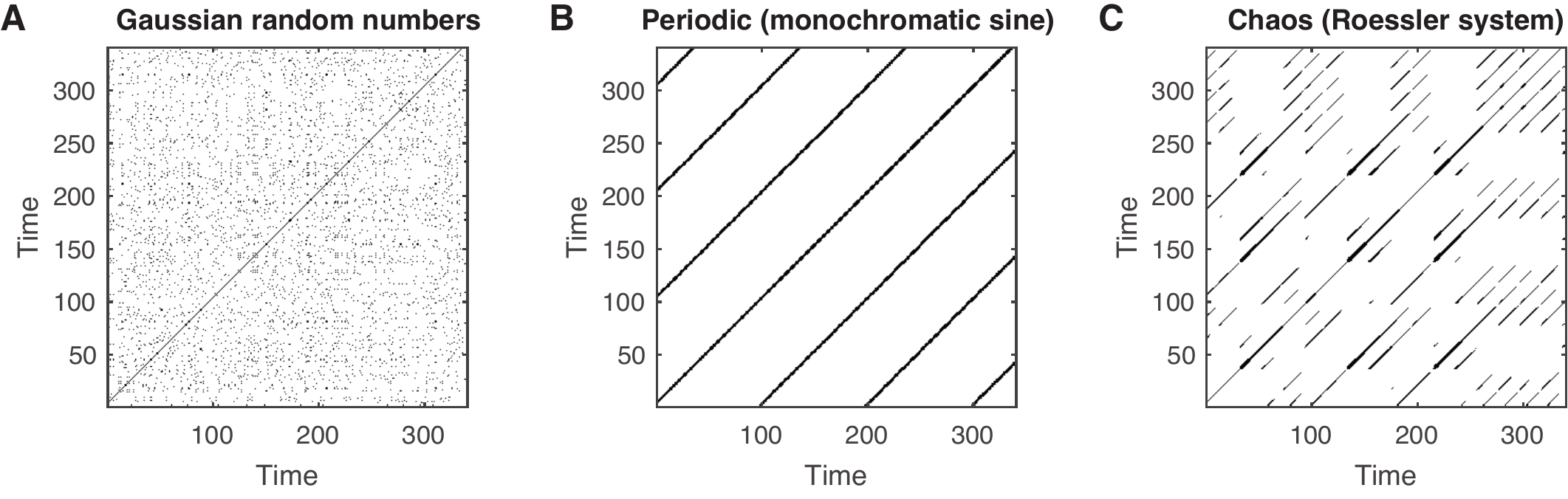}}
\caption{RPs of (A) standard normal Gaussian numbers, (B) time-delay embedded
sinusoidal with an oscillation period $T=100$ time units ($m=2$, $\tau=T/4$), and (C) the R\"ossler
system ($a=0.15$, $b=0.2$, $c=10$) (only subsets shown). RPs were constructed from time series of
2,000 samples (in case of the R\"ossler system we removed transients) using a constant global
recurrence rate of 4\% with a fixed threshold and Euclidean norm.}
\label{fig_example_rp}
\end{figure}

\begin{figure}
\centerline{\includegraphics[width=\textwidth]{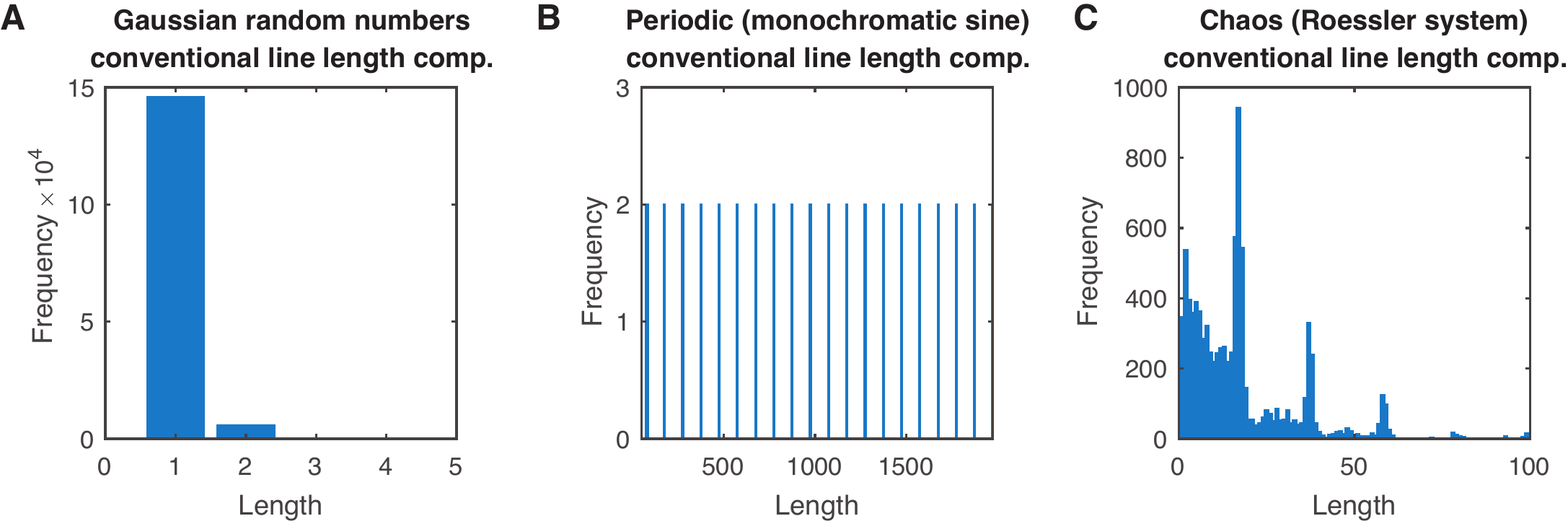}}
\caption{Diagonal line length distributions of the different systems types described in 
Fig.~\ref{fig_example_rp}, gained from the conventional line counting.}
\label{fig_example_hist_conventional}
\end{figure}

A recurrence plot (RP) is a binary, square matrix $\mathbf{R}$ representing the recurrences of states $\vec{x}_i$ ($i=1,...,N$, with $N$ 
the number of measurement points) in 
the $d$-dimensional phase space \cite{eckmann87,marwan2007}
\begin{equation}\label{eq_rp_definition}
{R}_{i,j}(\varepsilon) = \Theta\left(\varepsilon - \| \vec{x}_i - \vec{x}_j\|\right), \qquad \vec{x} \in \mathbb{R}^d,
\end{equation}
with $\|\cdot \|$ a norm, $\varepsilon$ a recurrence threshold, and $\Theta$ the
Heaviside function. The RP consists of small-scale structures, such as
single points and diagonal and vertical lines, which characterize important dynamical
properties of the system.
A diagonal line is a sequence of pairs of time points 
$\mathcal{L} \colonequals \{(i,j), (i+1,j+1),\ldots,(i+\ell-1,j+\ell-1)\}$
where ${R}_{i,j}  \equiv 1$ for all index pairs in $\mathcal{L}$.
Diagonal lines in the RP represent the temporal duration that 
two distinct parts of the phase space trajectory 
run parallel (Figs.~\ref{fig_eps-tube} and \ref{fig_example_rp}). The histogram $P(\ell)$ of the lengths 
of diagonal lines (Fig.~\ref{fig_example_hist_conventional}) characterizes the dynamics
\cite{faure98,thiel2004a,march2005}
and can be and has been
used to quantitatively distinguish between RPs, the underlying dynamics,
or to identify regime transitions \cite{trulla96,marwan2002herz,kopacek2010,mitra2014,nair2014}.

For uncorrelated noise, the probability to find a line $\mathcal{L}$ of exact length
$\ell$ decays exponentially \cite{thiel2003} (Fig.~\ref{fig_example_hist_conventional}A), i.e., the RP consists only of 
very short diagonal lines, if there are any lines at all (Fig.~\ref{fig_example_rp}A).
In contrast, for chaotic dynamics, 
the RP contains diagonal lines of different lengths (Fig.~\ref{fig_example_rp}C),
resulting in a broad distribution $P(\ell)$ (Fig.~\ref{fig_example_hist_conventional}C).
The RP for a periodic system
contains continuous, non-interrupted
diagonal lines, virtually of infinite length (Fig.~\ref{fig_example_rp}B). 
In principal, we would expect a discrete line length distribution with a peak at 
line length infinity.
However, the lines are cut at the begin and end of the RP, such that an uncorrected
conventional line length measurement results in
a discrete distribution $P(\ell)$ with uniform characteristics 
(Fig.~\ref{fig_example_hist_conventional}B).

The RP is a discrete matrix. Therefore, the creation of the histogram $P(\ell)$
appears to be trivial. But it is not as simple as it looks at the first
glance. Diagonal lines can be quite long and -- as already mentioned -- can exceed the finite
size of the RP. In practice, this is a very common problem, particularly when
a sliding window method is applied. How to count such diagonal lines? As we will
see later, for some measures, it can be important to have the correct length of 
the lines, for other measures it does not play any role. In the original 
definition, the lines are also counted even if they were cut by the RP border \cite{zbilut92,webber94,marwan2007}.

Several measures for RP analysis have been introduced which use $P(\ell)$.
The firstly introduced measure was the
determinism \cite{webber94}. This measure is the fraction of recurrence points that 
form diagonal lines
\begin{equation}\label{eq_det}
D = \frac{\sum_{\ell=\ell_{\min}}^N P(\ell)}{ \sum_{\ell=1}^{N} \ell P(\ell)}
\end{equation}
and considers lines which have at least length $\ell_{\min}$, which in principle is
a free parameter, but often set to 2. Nevertheless the choice of the minimal line length
can be crucial for the correct estimation of some RQA measures and we come back to that in
Sect.~\ref{sec_results_noise}. More
details about this can be found in \cite{marwan2007}.
Since RPs of uncorrelated noise have mainly single points and only
few and short diagonal lines, for such dynamics $D$ has rather low
values (although embedding can result in artificially high $D$ values, see
discussion in \cite{marwan2011,wendi2018}). In contrast, RPs for deterministic
dynamics contain of many diagonal lines, resulting in elevated values of $D$,
with the special case of $D = 1$ for periodic and quasi-periodic dynamics.
As this measure only quantifies whether a recurrence point is on 
a diagonal line or not, the actual length of a diagonal line is not
important (i.e., whether the line crosses the RP border or not).

Another idea is to look at the average and maximal length of the detected
diagonal lines (related to prediction time and Lyapunov exponent, resp.\cite{marwan2007}). 
The average, of course, depends on the actual line lengths and will
be biased when diagonal lines cross the RP borders.

Because the shape of $P(\ell)$ differs for different dynamics, the Shannon
entropy of the probability distribution $p(\ell) = \frac{P(\ell)}{\sum_{\ell} P(\ell)}$
to find a diagonal line of exact length $\ell$ was suggested \cite{webber94}
\begin{equation}\label{eq_S_RQA}
S = -\sum_{\ell=\ell_{\min}}^{N} p(\ell) \ln p(\ell).
\end{equation}
This measure was introduced in a pragmatic way to quantify the visual
line structures in the RP and has been interpreted as the ``information content of
the trajectories'' \cite{zbilut98a}.
Here, the choice of the minimal line length $\ell_{\min}$ has a significant 
effect, since it discards parts of the line length histogram and
therefore alters its shape.
For uncorrelated noise, $S$ has low values, because $p(\ell)$ is exponentially
decaying. For chaotic dynamics, $p(\ell)$ is a broad distribution, resulting
in quite large $S$ values. However, for periodic signals $p(\ell)$ has more similarity
with a uniform distribution if the mentioned border effects are not accounted for.
Therefore, $S$ is not low for periodic signals, although we would expect it,
but rather large, even larger than for chaotic dynamics.
Here, the effect of the sliced lines at the RP border has the strongest
and remarkable effect, which is why we focus on this measure only in this letter.
Maximal and mean diagonal line length and specifically determinism and 
their behavior with respect to border effects, the choice of further RP related parameters and
its interpretation will be examined carefully in a forthcoming paper.

\section{Correction schemes for counting diagonal lines}\label{sec_correction_schemes}

In this section we show two ways of overcoming the problem of biased diagonal line based measures
due to the border effect. Either we manipulate the histogram of the
diagonal lines (Sect.~\ref{sec_linecount_correction}) or we change the shape of the RP in order
to avoid a bias in the first place (Sect.~\ref{sec_windowshapes_correction}).

\subsection{Alternative ways of counting line lengths}\label{sec_linecount_correction}

Let $\textbf{R}$ be a $N\times N$ recurrence matrix, Eq.~\eqref{eq_rp_definition}, and $P(\ell)$ the
histogram of the diagonal lines contained in $\textbf{R}$. 
We now substantiate the definition of a diagonal
line in an RP from Sect.~\ref{sec_rqa_border}. A diagonal line $\mathcal{L}$ of length $\ell$ is a set 
of $\ell$ index tupels $(\cdot,\cdot)_{k=1,..,\ell}$:
\begin{equation}\label{eq_diagonal_line}
\mathcal{L}_{\ell} \colonequals 
	\lbrace (i+k,j+k) \ |
	\forall k=0,...,\ell-1\ :\ 
	\left( 1 - {R}_{i-1,j-1} \right)	
	\left( 1 -{R}_{i+\ell,j+\ell} \right)
	{R}_{i+k,j+k} \equiv 1 \rbrace.
\end{equation} 
The length of a line $\ell,$ is usually the cardinality of this set $|\mathcal{L}|$.

We denote any diagonal line which
starts and ends at the border of $\textbf{R}$ as a \textit{border diagonal},
e.g., in case of the lower triangle of the RP, when starting at $(i,1)$ in the first column
and ending at $(N,N-i+1)$ in the last row:
\begin{align}\label{eq_border_line}
\mathcal{L}_\text{border} \colonequals 
&\lbrace (i+k,1+k) \ | 
	\ \forall\, k=0,\ldots,N-i\ :\ {R}_{i+k,1+k} \equiv 1 \quad \vee \notag\\
& (1+k,j+k) \ |
	\ \forall \, k=0,\ldots,N-j\ :\ {R}_{1+k,j+k} \equiv 1\rbrace.
\end{align} 
Any diagonal of length $\ell$, which starts or ends at the border of $\textbf{R}$ and 
has an end or start point within the recurrence matrix, we call
\textit{semi border diagonal}:
\begin{align}\label{eq_semi_border_line}
\mathcal{L}_\text{semi border} \colonequals 
\lbrace &(i+k,j+k) \ |
	\ \forall k=0, \ldots ,\ell-1 \wedge (j=1 \vee i + \ell-1 = N)\ :\ 
	{R}_{i+k,j+k} \equiv 1 \quad \vee \notag\\
& (i+k,j+k)\ |
	\ \forall k=0, \ldots ,\ell-1 \wedge (i=1 \vee j + \ell-1 = N)\ :\ 
	{R}_{i+k,j+k} \equiv 1\rbrace.
\end{align} 

\subsubsection{Discard border diagonals from histogram (dibo correction)}\label{sec_dl2_correction}

The real length of the border diagonals is unknown. Therefore, 
we are not able to assign their true length to them and, hence, one option to 
deal with the missing length regarding the line length histogram is 
setting their length to zero. That is, we simply discard all (semi-)border diagonals from $P(\ell)$ and, thus,
avoid the broad line length distribution as exemplary shown in Fig.~\ref{fig_example_hist_conventional}B.
As desired, this results in a lowered entropy value, but also has some
drawbacks. In case of a
perfectly sampled stationary periodic signal (without any noise contamination and
without effects due to tangential motion, cf. Sect.\ref{sec_tangential_motion}) 
this method would
empty the histogram $P(\ell)$ completely, leaving an undefined entropy 
(Figs.~\ref{fig_results_1}, 
\ref{fig_appendix_results_1}, \ref{fig_appendix_results_2})
and a mean and maximum line length of zero
(cf. Fig.~\ref{fig_correction_histograms}B for a result not corrected for tangential motion).
In the following, we refer to this approach as {\it dibo correction} (DIscard
BOrder diagonals).

\subsubsection{Assign maximum line length to all border diagonals (Censi correction)}\label{sec_censi_correction}

\begin{figure}
\begin{center}
\includegraphics[width=1\textwidth]{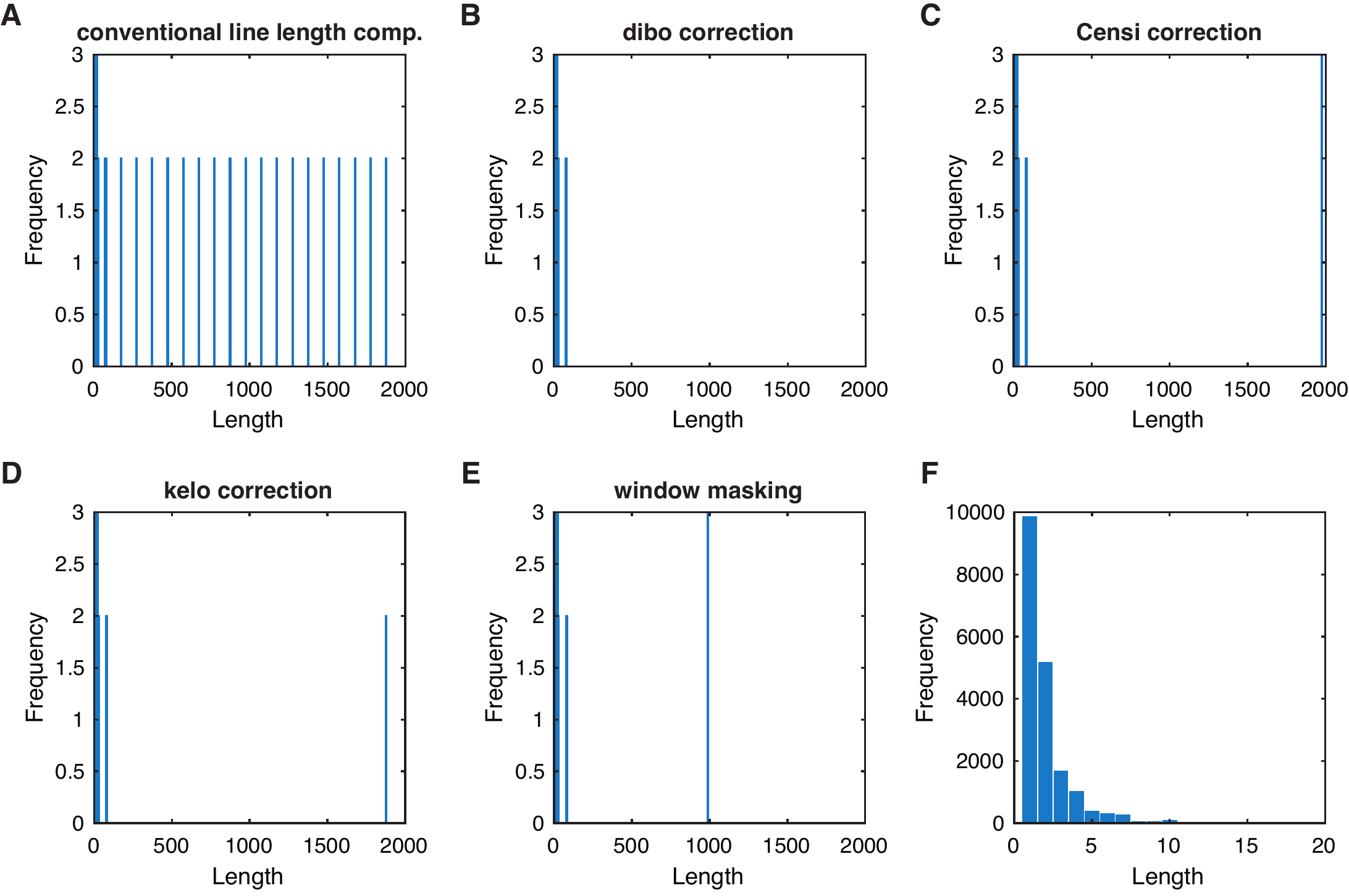}

\caption{Diagonal line length histograms of the conventional line length computation (A) and
of the presented correction schemes (B-E) for a monochromatic
time-delay embedded sinusoidal with an oscillation period $T=100$ time units 
($m=2$, $\tau=T/4$, same as in Figs.~\ref{fig_example_rp}B and \ref{fig_example_hist_conventional}B).
An enlargement of the histograms from panels A to D, focusing on the shorter line lengths, is presented
in panel F. A corresponding enlargement of panel E does qualitatively look the same, but with reduced
frequencies, due to the smaller effective window size (see text for details). 
For a better visibility we enlarged single bars in panels B to E and limited the view
to a frequency range $[0~3]$ in panels A to E (in F the full range is used).
}\label{fig_correction_histograms}
\end{center}
\end{figure}

To avoid an empty diagonal line histogram, Censi et al.\cite{censi2004} suggested to assign 
all border diagonals the length of the main diagonal of the RP (line of identity). Sticking to the
aforementioned example of a
perfectly sampled and uncontaminated stationary periodic signal, this
modification would result in
a delta peak in $P(\ell)$ (cf. Fig.~\ref{fig_correction_histograms}C for a result not corrected for tangential motion), and therefore a sound defined
entropy value of zero as well
as meaningful mean and maximal line length estimate
(Figs.~\ref{fig_results_1}, 
\ref{fig_appendix_results_1}, \ref{fig_appendix_results_2}).
For deterministic chaotic processes this correction scheme
could underestimate the entropy, if 
the RP is smaller than the average length scale of the diagonal lines. Especially in a running window
approach, this effect is assumed to be significant.
We refer to this approach as {\it Censi correction}.

\subsubsection{Keeping just the longest border diagonal (kelo correction)}\label{sec_dl3_correction}

In alternative to the correction in Sect.~\ref{sec_dl2_correction}, all 
(semi-)border diagonals from $P(\ell)$ are discarded, but the longest one
(cf. Fig.~\ref{fig_correction_histograms}D). This approach would also avoid the 
broad line length
distribution shown in Figs.~\ref{fig_example_hist_conventional}B and 
~\ref{fig_correction_histograms}A, 
but would leave a valid definition of the entropy, since $P(\ell)$ is 
not an empty set (cf. Figs.~\ref{fig_results_1}, 
\ref{fig_appendix_results_1}, \ref{fig_appendix_results_2}). 
The resulting entropy for the aforementioned example would be low. 
In contrast to the Censi correction, this approach would avoid the bias
for deterministic chaotic processes when a windowing approach is applied.
In the following, we refer to this approach as {\it kelo correction} (KEeping
the LOngest border diagonal).

\subsection{Alternative RP window shapes (window masking)}\label{sec_windowshapes_correction}

\begin{figure}
\begin{center}
\includegraphics[width=0.33\textwidth]{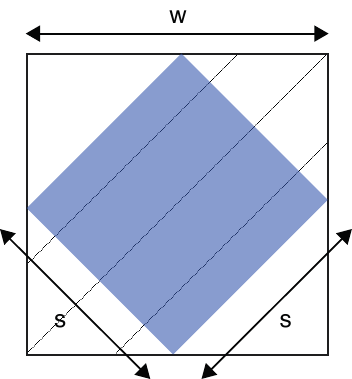}

\caption{Blue shaded alternative window shape with edge length $s$ of a $w\times w$ recurrence plot. 
$s$ and $w$ imply the number of RP matrix elements covered by the window shapes.}
\label{fig_windowshape3}
\end{center}
\end{figure}

The origin of the border diagonals is related to a geometric difference between
the RP and the diagonals. Therefore, a further approach to avoid the length bias
of border diagonals is to apply a specific window to the RP which has the 
same geometric orientation as the diagonals. 
One realization of such a window is a {45\textdegree} rotated cutout from the
original RP (Fig.~\ref{fig_windowshape3}). 
Conventionally counting the lines of this rotated RP cutout
preserves a delta peak distribution in $P(\ell)$ in case of a periodic signal 
(cf. Fig.~\ref{fig_correction_histograms}E for a result not corrected for tangential
motion).
However, with this shape we loose $w^2 - 2s^2 = w^2 - \frac{1}{2}w^2 = \frac{1}{2}w^2$ data points
with respect to the original RP. Note that $s$ and $w$ in Fig.~\ref{fig_windowshape3} imply a number of
data points, meaning hypotenuse and catheti of an isosceles triangle have the same length 
($w=\frac{s}{2}$).
We argue that this approach could be rather useful in a running window
approach over a global RP, where the size of the alternative shape could be chosen such that 
it contains as many data points as the \textit{classic}, non-rotated, window.
We refer to this approach as {\it window masking}.
An alternative window shape would be a parallelogram with the top and bottom sides having the
{45\textdegree} direction \cite{donath2016}.

\section{Tangential motion in Recurrence Quantification Analysis}
\label{sec_tangential_motion}

Even though the considerations made in the preceding sections are valid and useful, the correction schemes
presented in Sect.~\ref{sec_correction_schemes} most likely do not give the expected correction
for the entropy of diagonal line lengths for experimental data, unless the data has been
properly preprocessed. There are three reasons why the correction of
the border diagonals in the diagonal line histogram $P(\ell)$ is not sufficient
enough: (\textit{i}) temporal
correlations in the data, especially when highly sampled flow data is used, (\textit{ii})
noise, and (\textit{iii})
insufficient embedding of the time series at hand (if needed) combined with the 
effect of discretization and an inadequate 
choice of parameters needed to construct the RP (recurrence threshold method, recurrence threshold size, norm). 

\begin{figure}
\begin{center}
\includegraphics[width=\textwidth]{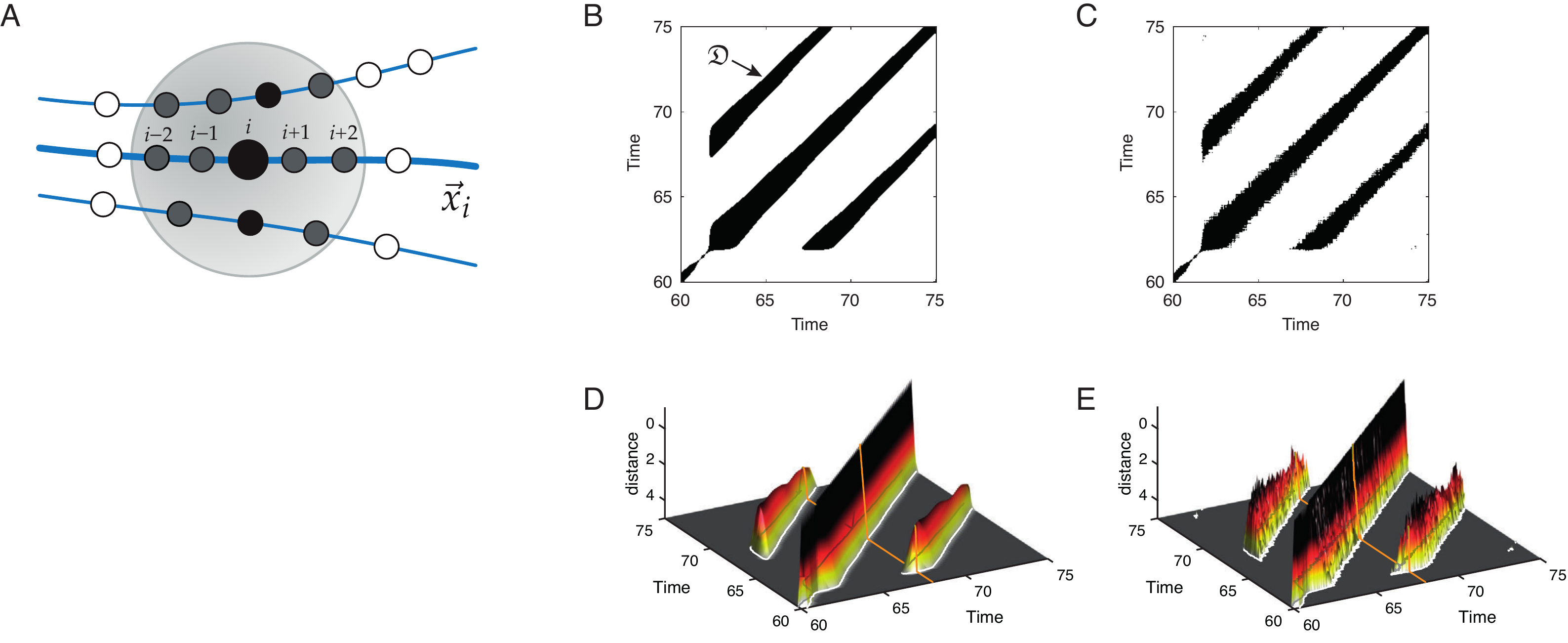}

\caption{(A) Tangential motion, i.e., points of a trajectory preceding and succeeding
a (recurring) state (gray), cause thickening of diagonal lines in the RP (B), (C).
The thickening of diagonal lines can vary, e.g., as in this example
of the R\"ossler system (noise free case in B and additive noise in C). The diagonal lines are more thick at the beginning
and become less thick with time. A diagonal line in an RP (B), (C) denotes a range of 
distances in the distance matrix falling under a the recurrence threshold $\varepsilon$.
Panels (D) and (E) show three ``distance ranges'' (we call such a range $\mathfrak{D}$ in the text)
corresponding to the three lines in (B), (C) respectively. Shown is a colorcoded, thresholded
distance matrix with reversed {$z$-axis} for a better visibility (increasing distances
from top to bottom). The colormap encodes zero distance as black and the distance
corresponding to the recurrence threshold as grey.}\label{fig_tangentialmotion}
\end{center}
\end{figure}

Temporal correlation means that states $\vec{x}_{j}$ preceding or succeeding a state 
$\vec{x}_{i}$ (or a recurring state $\vec{x}_{k}$ of $\vec{x}_{i}$), are very similar 
to this one and, hence, falling into the 
neighbourhood of $\vec{x}_i$ (or $\vec{x}_k$) and to be considered as recurrences, 
i.e., $R_{i,j} \colonequals 1$ for $j = [i-m,\ldots,i+n]$ or
for $j = [k-m,\ldots,k+n]$ when $R_{i,k} \colonequals 1$
(Fig.~\ref{fig_tangentialmotion}A). This results in 
vertically extended sequences in the RP, i.e., thickening its diagonal lines.
The thickening leads to an artificially enlarged number of diagonal lines, thus effecting
the distribution $P(\ell)$, and is
often referred to as \textit{tangential motion} \cite{gao2000, marwan2007}.
Moreover, the thickening is not evenly distributed along a diagonal line 
(Fig.~\ref{fig_tangentialmotion}B). For border diagonals, this means that 
there are not only additional border diagonals (which could be handled by applying
correction schemes as described in Sect.~\ref{sec_correction_schemes}), but
additional shorter diagonal lines, again leading to a broadening of the 
line length distribution $P(\ell)$ (Fig.~\ref{fig_correction_histograms}, in
particular panel F) 
and an elevated entropy $S$. 

Additive noise causes the already thickened lines in the RP to appear more diffus (Fig.~\ref{fig_tangentialmotion}C,E). Technically
speaking, the noise alters the phase space trajectory, causing the pairwise distances to randomly
scatter about their true/noise free values and, thus, the histogram $P(\ell)$ gets
enriched with small line lengths \cite{thiel2002}. 
This eventually biases the RQA measures discussed in Sect.~\ref{sec_rqa_border}. 

\section{Correction schemes for reducing the effects of tangential motion}\label{sec_correction_schemes_tangmotion}

\subsection{Perpendicular RP}\label{sec_perp_rp}

A straightforward way to reduce the thickening of the diagonal lines
from a theoretical
perspective is the \textit{perpendicular RP}, suggested by Choi et al. \cite{choi99}

\begin{equation}\label{eq_perpendicular_RP}
\textbf{R}_{i,j}^{\perp}(\varepsilon) = \Theta\left(\varepsilon - \| \vec{x}_i - \vec{x}_j\|\right)\cdot \delta\left(\dot{\vec{x}}_i \cdot (\vec{x}_i - \vec{x}_j)\right),
\qquad \vec{x} \in \mathbb{R}^d.
\end{equation}

This RP contains only those points $\vec{x}_j$ that fall into the neighbourhood of $\vec{x}_i$ and lie in the ($d-1$)-dimensional
subspace of $\mathbb{R}^d$ that is perpendicular to the phase space trajectory at $\vec{x}_i$. Although theoretically there is no
need for an additional parameter in order to construct a perpendicular RP, 
in practical situations
almost no points in $\mathbb{R}^d$ phase space end up on the mentioned ($d-1$)-dimensional subspace of $\vec{x}_i$
(Poincar\'{e} section), due to limited resolution (discretization) of the data. Hence, it is reasonable to introduce an additional threshold parameter 
$\varphi$, which allows points $\vec{x}_j$ to be considered as perpendicular to $\vec{x}_i$, if 

\begin{equation}
\arccos{\frac{\dot{\vec{x}}_i \cdot (\vec{x}_i-\vec{x}_j)}{|\dot{\vec{x}}_i| \cdot |(\vec{x}_i-\vec{x}_j)|}} 
\in \Bigl[\left(\frac{\pi}{2}-\varphi\right),
\left(\frac{\pi}{2}+\varphi\right)\Bigr].
\end{equation}

Thus, Eq.~\eqref{eq_perpendicular_RP} transforms to

\begin{equation}\label{eq_perpendicular_RP_2}
{R}_{i,j}^{\perp}(\varepsilon,\varphi) = \Theta\left(\varepsilon - \| \vec{x}_i - \vec{x}_j\|\right) \cdot
\Theta\left( \varphi - \biggl| \arccos{\frac{\dot{\vec{x}}_i \cdot (\vec{x}_i-\vec{x}_j)}{|\dot{\vec{x}}_i| \cdot |(\vec{x}_i-\vec{x}_j)|}}\biggl| - \frac{\pi}{2}  \right),
\qquad \vec{x} \in \mathbb{R}^d.
\end{equation}

Figure \ref{fig_rp_corrections}B shows a perpendicular RP for a R\"ossler system (with parameters $a=0.15$, $b=0.2$, $c=10$,
transients removed). For the estimation of the tangential at each point in phase space we used the reference point, its predecessor 
and its successor. We set the angle threshold to $\varphi = \frac{\pi}{12}$ ($=15$\textdegree).

\subsection{Isodirectional RP}\label{sec_iso_rp}

\begin{figure}
\begin{center}
\includegraphics[width=\textwidth]{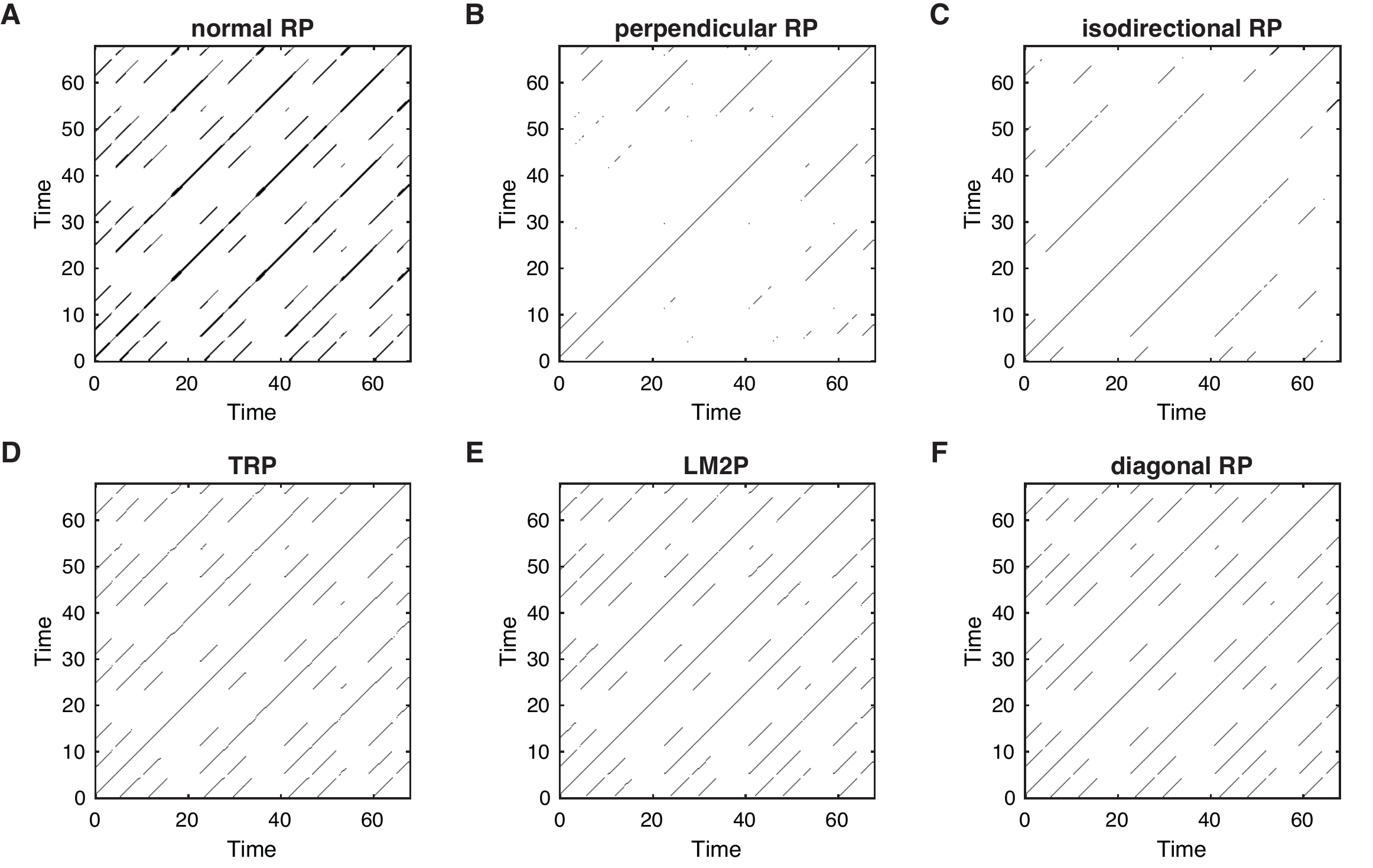}

\caption{Different approaches for avoiding the effect of tangential motion in a recurrence plot (RP), exemplary shown for the R\"ossler
system (with parameters $a=0.15$, $b=0.2$, $c=10$, sampling time $\Delta t=0.2$). 
(A) Normal RP with fixed recurrence threshold ensuring
4\% global recurrence rate as a basis to all other RPs shown in this figure. 
(B) Perpendicular RP with angle threshold $\varphi = 15$\textdegree,
(C) isodirectional RP with $T=5$ [sampling units] and $\varepsilon_2 = \varepsilon/2$, 
(D) true recurrence point RP (TRP) with $T_\text{min}=5$ [sampling units], 
which coincides with the first minimum of the mutual information,
(E) thresholded local minima approach with two parameters (LM2P) and $\tau_\text{m}=5$, and
(F) diagonal RP.}\label{fig_rp_corrections}
\end{center}
\end{figure}

Requiring less computational effort, the \textit{iso-directional RP} suggested by
Horai et al. \cite{horai2002} also promises to cope with the tangential motion, but also
inherits two additional parameters $T$ and $\varepsilon_2$ (Fig.~\ref{fig_rp_corrections}C). In this
approach two points in phase space are denoted recurrent, if their mutual distance falls within the
recurrence threshold $\varepsilon$ and the distance of their trajectories
throughout $T$ consecutive time steps falls within a recurrence threshold $\varepsilon_2$

\begin{equation}\label{eq_iso_RP}
{R}_{i,j}^{\Nearrow}(\varepsilon,\varepsilon_2,T) = \Theta(\varepsilon - \| \vec{x}_i - \vec{x}_j\|) \cdot
\Theta(\varepsilon_2 - \| (\vec{x}_{i+T} - \vec{x}_i) - (\vec{x}_{j+T} - \vec{x}_j)\|),
\qquad \vec{x} \in \mathbb{R}^d.
\end{equation}

We achieved decent results when choosing $T$ in the size of the decorrelation time (e.g., first minimum
of the mutual information) and the second recurrence threshold as half of the size of the recurrence
threshold $\varepsilon$, which determines the parent RP.

\subsection{True recurrence point RP (TRP)}\label{sec_trp}

Inspired by the work of Gao \cite{gao99}, Ahlstrom et al. \cite{ahlstrom2006b} compute a normal RP, 
Eq.~\eqref{eq_rp_definition}, but only accept those
points to be recurrent, which ``first'' enter the $\varepsilon$ neighbourhood shown in 
Fig.~\ref{fig_tangentialmotion}A.
To ensure this, they first identify all points which fall into an $\varepsilon$-neighbourhood of
a certain point $\vec{x}_i$ 

\begin{equation}
\zeta_i \equiv \{ \vec{x}_{j_1},\vec{x}_{j_2},... ~ |{R}_{i,j_{k}} \colonequals 1 \}, 
\end{equation}

i.e., all points $j_k$ in column $i$ of the RP. The time difference of two consecutive
recurrence points $\vec{x}_{j_k},\vec{x}_{j_{k+1}}$ is 
$\{T_k^{(1)} = j_{k+1} - j_k \}_{k \in \mathbb{N}}$ in units of the sampling
time (recurrence times of first type \cite{gao99}) and these correspond to the vertical distances 
between these points in column $i$ of the RP. They now 
discard all points from the RP, which vertical distance to its neighbouring point in a column is 1
and leaving all points with recurrence time larger than 1,

\begin{equation}\label{eq_ahlstrom_correct}
\zeta_i^* \equiv \{\vec{x}_{j_1},\vec{x}_{j_2},\ldots ~ |{R}_{i,j_{k}} \colonequals 1 , ~ T_k^{(1)} > T_\text{min} \}, \qquad T_\text{min} = 1.
\end{equation} 

The authors call this modified RP a \textit{true recurrence point recurrence plot (TRP)}. This is different than 
simply discarding all points from the computations of Eq.~\eqref{eq_rp_definition} 
which fall within a certain time range $w_\text{Theiler}$ of the reference point (\textit{Theiler window} 
\cite{theiler1986})

\begin{equation}\label{eq_rp_theiler}
{R}_{i,j}(\varepsilon) = \Theta\left(\varepsilon - \| \vec{x}_i - \vec{x}_j\|\right), \qquad |i-j|>w_\text{Theiler}, ~ \vec{x} \in \mathbb{R}^d.
\end{equation}

To obtain a TRP, we suggest to discard all recurrence points with recurrence times greater than $w_\text{Theiler}$, i.e.,
$T_\text{min} = w_\text{Theiler}$ in Eq.~\eqref{eq_ahlstrom_correct}. The Theiler window 
should be set in the order of the decorrelation time or the delay, if time delay
embedding is used for reconstructing the phase space vectors from time series. 

However, the TRP most often leads to disjoint, deviated diagonal line structures (Fig.~\ref{fig_rp_corrections}D),
which correspond to the white embraced lines in Fig.~\ref{fig_tangentialmotion}D, E.

An alternative would be to use the mid-points of the recurrence sequences. This would also
correspond to recurrence times as discussed in \cite{ngamga2012}. 
In Subsec.~\ref{sec_diagonal_RP}, we will develop another correction scheme which
is motivated by these mid-point based ``true recurrences''.

\subsection{RP by means of local minima}\label{sec_local_minima}

Another approach for reducing the effect of tangential motion and which shares
the basic idea from the TRP approach, was introduced by
Schultz et al. \cite{schultz2011}, who track the local minima of the distance matrix 
(corresponding to the maxima in Fig.~\ref{fig_tangentialmotion}D, E). Wendi \& Marwan \cite{wendi2018b} then extended this idea 
in order to make the method more robust against noise. However, such local-minima based RP can contain
bended or disrupted diagonal line structures.
The key idea is to look for local minima in each column of the distance matrix, illustrated as an orange cross section
in Fig.~\ref{fig_tangentialmotion}D, E. If such a local
minimum is smaller than the recurrence threshold, then it is a recurrence (LocalMinimaThresholded, LMT). In the
two-parameter approach (LM2P) \cite{wendi2018b} shown in Fig.~\ref{fig_rp_corrections}E, there is an additional constraint
for two consecutive local minima to be displaced by at least $\tau_\text{m}$ time steps.

\subsection{Diagonal RP}\label{sec_diagonal_RP}
  
We now propose an additional approach to cope with the tangential motion, which does not need
any additional parameters and leads to an RP of straight, unbended diagonal line structures (Fig.~\ref{fig_rp_corrections}F). 
We call this approach the \textit{diagonal RP}, since it generates an RP with only diagonal line structures that are just one point thick.

A diagonal line in a RP corresponds to a connected region in the distance matrix
with distances smaller than the recurrence threshold $\varepsilon$ (Fig.~\ref{fig_tangentialmotion}D, E, white embraced region). 
We call such
region a ``distance range'' $\mathfrak{D}$. Typically, the larger $\varepsilon$ the
larger the $\mathfrak{D}_i$'s in the RP. Moreover, tangential motion, noise and
insufficient embedding affect the shape and width of the $\mathfrak{D}_i$'s.
For the diagonal line based RQA measures we are
interested in these ranges to be represented by single, connected diagonal lines in the 
corresponding RP. We choose the longest line of each $\mathfrak{D}_i$ to be its
adequate representative in the RP. 
We define the ``distance ranges'' $\mathfrak{D}_i$ of an RP recursively as a set of 
adjacent diagonal lines $\mathcal{L}^{(m)}_{\ell_m}$ of length $\ell_m$ (cf. Eq.~\eqref{eq_diagonal_line}),
initializing with the longest line $\mathcal{L}^{(k)}_{\ell_k}$, for which $\ell_k = {\max}(\ell\ :\ P(\ell)>0)$.

\begin{align}\label{eq_definition_distance_range}
\mathfrak{D}_i \colonequals \lbrace \mathcal{L}^{(k)}_{\ell_k}, \mathcal{L}_{\ell_m}^{(m)} ~ | ~ &
\mathcal{L}_{\ell_m}^{(m)} ~ \curvearrowleft ~
\mathcal{L}_{\ell_{m-1}}^{(m-1)} ~ \curvearrowleft ~ \mathcal{L}_{\ell_{m-2}}^{(m-2)} 
~ \curvearrowleft ~ ... ~\curvearrowleft ~\mathcal{L}^{(k)}_{\ell_k} ~ \vee ~
\notag \\
& \mathcal{L}_{\ell_m}^{(m)} ~ \curvearrowright ~
\mathcal{L}_{\ell_{m-1}}^{(m-1)} ~ \curvearrowright ~ \mathcal{L}_{\ell_{m-2}}^{(m-2)} 
~ \curvearrowright ~ ... ~\curvearrowright ~\mathcal{L}^{(k)}_{\ell_k}
\rbrace  
\end{align} 

with the line-neighbour-relations $\curvearrowleft$ and $\curvearrowright$ defined by

\begin{align}\label{eq_definition_line_neighbourhood}
\exists p \in [1,...,\ell_m]~ \exists q \in [1,...,\ell_k]: \qquad\qquad\qquad\qquad\qquad\qquad\qquad\qquad  \notag \\
(i_m,j_m)_p \colonequals
\begin{cases}
    (i_k+1,j_k)_q \vee (i_k,j_k+1)_q, & \text{if}~\mathcal{L}_{\ell_m}^{(m)} \curvearrowleft \mathcal{L}_{\ell_k}^{(k)}\\
    (i_k-1,j_k)_q \vee (i_k,j_k-1)_q, & \text{if}~\mathcal{L}_{\ell_m}^{(m)} \curvearrowright \mathcal{L}_{\ell_k}^{(k)}
  \end{cases}
\end{align}

where $(i_m,j_m)_{p=[1,...,\ell_m]}$ denote the index tuples corresponding to lines $\mathcal{L}_{\ell_m}^{(m)}$ and
$(i_k,j_k)_{q=[1,...,\ell_k]}$ denote the index tuples corresponding to the longest line $\mathcal{L}_{\ell_k}^{(k)}$. We then 
delete all lines contained in $\mathfrak{D}_{i}$ from the histogram $P(\ell)$ and define the next distance range 
$\mathfrak{D}_{i+1}$ with a new $\mathcal{L}_{\ell_{k'}}^{(k')}$ from the histogram and so on until $P(\ell)$ is an empty set.

We construct the \textit{new} RP by keeping the longest line of each $\mathfrak{D}_i$ (all the $\mathcal{L}_{\ell_k}^{(k)}$'s).
Denote the set of index tuples $(i,j)$ corresponding to the set of longest lines gained from the $\mathfrak{D}_i$'s
as $\mathfrak{S}$, then
 
\begin{equation}
{R}^{\nearrow}_{i,j} = 
\begin{cases}
1, ~ \text{if}~ (i,j) \in \mathfrak{S} \\
0, ~ \text{otherwise}
\end{cases}
\end{equation}

Note that this algorithm constricts clusters of adjacent recurrence points to a single diagonal line, representing this 
``distance range'' $\mathfrak{D}$ (skeletonization). Although this method impresses with the absence of additional parameters, caution in
its use is advised concerning the choice of the embedding parameters and the recurrence threshold. A wrong setup, specifically
a too high recurrence threshold and/or a ``wrong'' time delay, could lead to an overall connected RP, which in turn would cause
a \textit{diagonal RP} consisting of just one single line in each triangle (if the main diagonal is discarded). However, concerning the
sensitivity to the choice of the recurrence threshold, our numerical investigations suggest a rather low risk of this special case and
a broad range of threshold values, which do work well (cf.~Sect.~\ref{sec_results}, Sect.~\ref{sec_appendix_sensitivity_threshold} and figures therein).

\section{Results: Efficience of correction schemes}\label{sec_results}

We now apply the correction schemes for counting diagonal lines (Sect.~\ref{sec_correction_schemes}) and
suppressing tangential motion (Sect.~\ref{sec_correction_schemes_tangmotion}) on a time discrete as well as 
time continuous example, in order to test their ability to give valid estimates for the entropy of
diagonal line lengths, Eq.~\eqref{eq_S_RQA}. In case of the former we choose the Logistic map $x_{n+1} = rx_n(1-x_n)$ with 
control parameter $r=3.5$, leading to regular limit cycle behavior, and control parameter $r=3.8$, 
where a chaotic regime is obtained. For the latter we show diagonal line length entropies of RPs of
the R\"ossler system \cite{roessler1976}

\begin{align}\label{eq_roessler_system}
\dot{x} &= -y -z \notag \\
\dot{y} &= x + ay \\
\dot{z} &= b + z(x-c) \notag
\end{align}

in two parameter configurations, also leading to regular limit cycle behavior ($a=0.15, b=c=10$) 
and chaotic motion ($a=0.15, b=0.2, c=10$) \cite{barrio2009}. The results shown 
in this section are based on ensembles of 100 realizations of each parameter setting for the
R\"ossler system and on ensembles of 1,000 realizations of each parameter setting for the Logistic map, 
gained from
randomly chosen initial conditions out of a uniformly distributed interval 
$x_0 \in [0, 0.5]$ (Logistic map), $x(0), y(0), z(0) \in [0, 2]$
(R\"ossler system). We numerically integrate the R\"ossler equations using 
the explicit Runge-Kutta (4,5) formalism (Dormand-Prince pair) as provided by 
the ode45-solver in MATLAB \cite{reichelt1994} with a fixed sampling time of $\Delta t=0.2$.
For both systems we discard the first 2,500 data points as transients, keeping 1,000 (Logistic map) 
and 2,000 (R\"ossler) data points as the time series we base our further
computations on. For estimating the entropy, we use the Maximum-Likelihood-estimator
$p(\ell)~ \widehat{=}~ \hat{p}(\ell) = \frac{\#\text{number of lines of length} ~ \ell}{\# \text{number of all lines in the RP}}$
for the probabilities. 

Generally, we expect (near-)zero entropy values for the regular regime setups and high(er) values for
the chaotic regime setups for both considered examples in the noise free
case (cf. Sect.~\ref{sec_rqa_border}). Moreover, we expect the correction schemes for counting diagonal 
lines (Sect.~\ref{sec_correction_schemes}) to perform well in case of the Logistic map examples, due to the
absence of tangential motion. For the flow data in the R\"ossler examples, we expect a combination of these
correction schemes with the correction schemes for tangential motion described in Sect. ~\ref{sec_correction_schemes_tangmotion}
to give reasonable results.
In order to validate our results we compute the diagonal line length entropy analytically for the mentioned cases.
March et al. \cite{march2005} gave an expression for this:

\begin{equation}
S_{\text{theoretical}} = K_2 \left( \frac{1}{\gamma} - 1 \right) - \ln{\gamma}\quad,
\label{eq_entropy_theo}
\end{equation}

with $\gamma = (1-e^{-K_2})$ and $K_2$ the correlation entropy. Practically we compute the largest Lyapunov exponent for
our experimental settings \cite{datseris2018} and use Pesin's identity to get the Kolmogorov entropy $K_1$. Because the correlation entropy is a lower
bound for the Kolmogorov entropy \cite{grassberger83a}, we expect the reference values computed from Eq.~\eqref{eq_entropy_theo} to give underestimated expectation values for the diagonal line length entropy.

\begin{figure}
\begin{center}
\includegraphics[width=.8\textwidth]{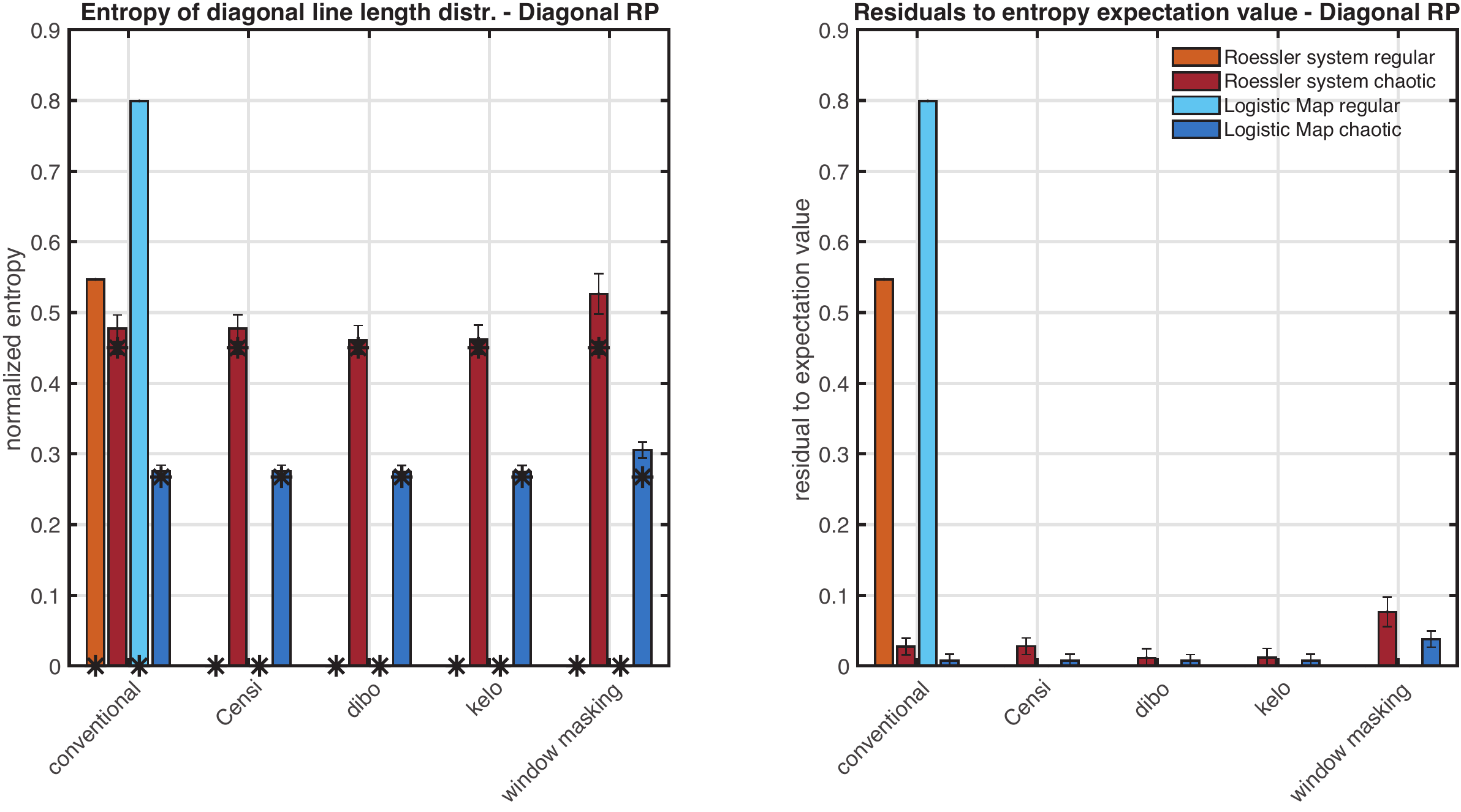}

\caption{Diagonal line length entropy (left panel) of the proposed diagonal recurrence plot ${R}^{\nearrow}$ 
(cf. Sect.~\ref{sec_diagonal_RP}) of the
R\"ossler system (reddish) and the Logistic map (bluish) in a regular limit cycle regime (bright) as well as in a chaotic regime
(dark). Shown are medians of the diagonal line length entropies gained from 1,000 realizations of the
Logistic map and 100 realizations of the R\"ossler example, respectively,
for the different line counting correction schemes described in Sect.~\ref{sec_correction_schemes}. 
Errorbars indicate two standard deviations of these distributions. Black stars show medians of ensembles of 1,000 analytically computed values derived from Eq.~\eqref{eq_entropy_theo} (its errorbars, as two standard deviations of
the ensemble distribution, are barely visible and smaller than
markers used). In the right panel the residuals to these underestimated expectation values are shown.
Firstly, RPs were obtained with a fixed recurrence threshold corresponding to 19\% recurrence rate in case of 
the R\"ossler examples and a fixed recurrence threshold corresponding to $1/10$ of the 
range of the underlying time
series in case of the Logistic map examples (for noise free map data the $\varepsilon$-adjustment with respect to the global
recurrence rate does not work properly). Then our proposed, parameter free correction scheme 
leading to the diagonal recurrence plot ${R}^{\nearrow}$ was applied. Results for a range of recurrence thresholds and for all
tangential motion RP-correction
schemes are shown in Fig.~\ref{fig_appendix_results_1} and Fig.~\ref{fig_appendix_results_2} in the Appendix
 \ref{sec_appendix_sensitivity_threshold}.}\label{fig_results_1}
\end{center}
\end{figure}

\begin{figure}
\begin{center}
\includegraphics[width=.8\textwidth]{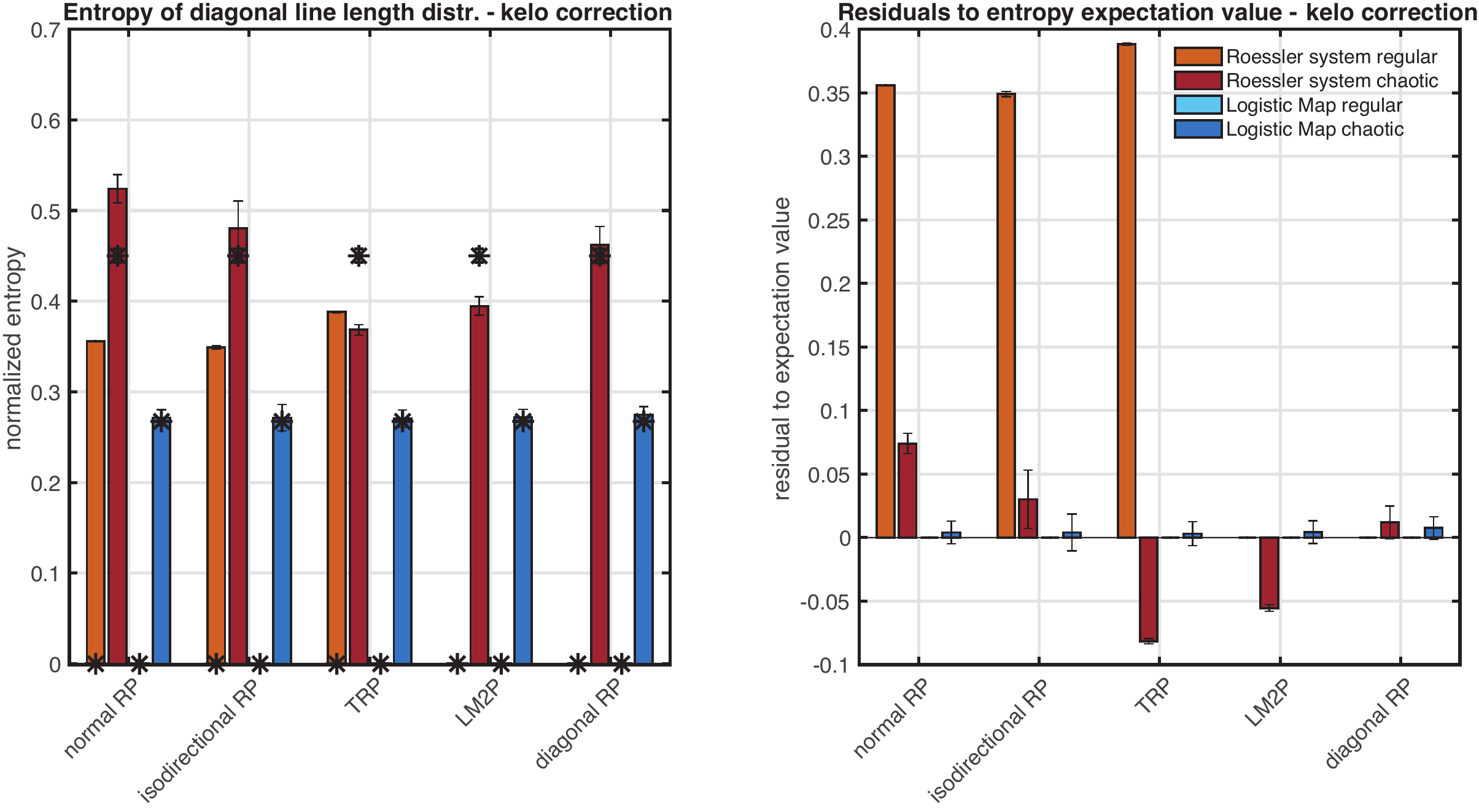}

\caption{Diagonal line length entropy (left panel) based on the proposed line counting correction scheme $kelo$
(cf. Sect.~\ref{sec_dl3_correction}) for the R\"ossler system (reddish) and the Logistic map (bluish) in a regular limit cycle 
regime (bright) as well as in a choatic regime (dark). 
Shown are medians of the diagonal line length entropies gained from 1,000 realizations of the
Logistic map and 100 realizations of the R\"ossler example, respectively, for
all the different tangential motion correction schemes described in Sect.~\ref{sec_tangential_motion}, but the perpendicular
recurrence plot ${R}^{\perp}$. 
Errorbars indicate two standard deviations of these distributions.
Black stars show medians of ensembles of 1,000 analytically computed values
derived from Eq.~\eqref{eq_entropy_theo} (its errorbars, as two standard deviations of
the ensemble distribution, are barely visible and smaller than
markers used). In the right panel the residuals to these underestimated expectation values are shown. 
The normal RP with a fixed recurrence threshold corresponding to 19\% recurrence rate in case of 
the R\"ossler examples and a fixed recurrence threshold corresponding to $1/10$ of the 
range of the underlying time
series in case of the Logistic map examples (for noise free map data the $\varepsilon$-adjustment with respect to the global
recurrence rate does not work properly) serves as a basis for the RP correction schemes shown here. Results for a range of 
recurrence thresholds and for all tangential motion RP-correction
schemes are shown in Fig.~\ref{fig_appendix_results_1} and Fig.~\ref{fig_appendix_results_2} 
in the Appendix \ref{sec_appendix_sensitivity_threshold}.}\label{fig_results_2}
\end{center}
\end{figure}

The results confirm our expectations (Fig.~\ref{fig_results_1}). 
While the conventional way of counting diagonal lines, where border effects are not taken into consideration, lead to
counterintuitive behavior, all the described correction schemes are able to distinguish chaotic from regular regimes in both
exemplary systems. In this laboratory, noise free conditions, the entropy estimates in case of the regular limit cycle regimes
are zero (or in case of the $dibo$-correction scheme not defined, due to the absence of any diagonal line). For $dibo$ and $kelo$
the estimated values for the chaotic R\"ossler regime fall within the two standard deviation margin of the theoretical values,
whereas Censi's correction scheme comes very close and the windowshape correction scheme misses it by approx.~5\%.
Again, we have to stress that we expect the expectation values to be underestimated, i.e. we assume Censi's method
and the window masking do also perform well. 
Let us stick to the $kelo$
correction scheme for now and look how the different correction schemes for tangential motion perform (Fig.~\ref{fig_results_2}).
First of all we have to mention that we were not able to produce any kind of reasonable estimates
while using the perpendicular recurrence plot ${R}^{\perp}$, regardless of the angle threshold parameter. This straightforward approach
is extremely sensitive to any kind of noise and to the sampling time of the system under observation. It needs a fairly high density of state 
space points, 
in order to yield a non empty RP and, thus, any meaningful diagonal line length entropy estimate. Hence, we skip this approach in our further analysis,
especially the dependence of the shown results to the choice of the recurrence
threshold and additive noise, but will discuss the performance of the perpendicular RP for a high sampled R\"ossler
setup in the next subsection. For a general use, we cannot recommend the application of 
perpendicular RPs. Coming back to the results (Fig.~\ref{fig_results_2}), solely the LM2P approach and the
diagonal RP perform as expected (zero-values in case of the regular regime setups and higher values for the chaotic regimes, clearly
distinguishable). Only the proposed diagonal RP is able to give estimates within the errorbars of the theoretical values (which is why
only this approach was selected for Fig.~\ref{fig_results_1}). Note that the reference values slightly underestimate the ``true'' value and we
cannot quantitatively correct for this bias.
As in Fig.~\ref{fig_rp_corrections}, we set the parameters $T$, $T_\text{min}$ and $\tau_\text{m}$ to the
corresponding first minimum of the auto mutual information and the second recurrence threshold for the isodirectional RP was again set
to $\varepsilon_2 = \varepsilon/2$, but we tried many parameter configurations.

\subsection{Results for high sampled data and the effect of noise}\label{sec_results_noise}

\begin{figure}
\begin{center}
\includegraphics[width=.99\textwidth]{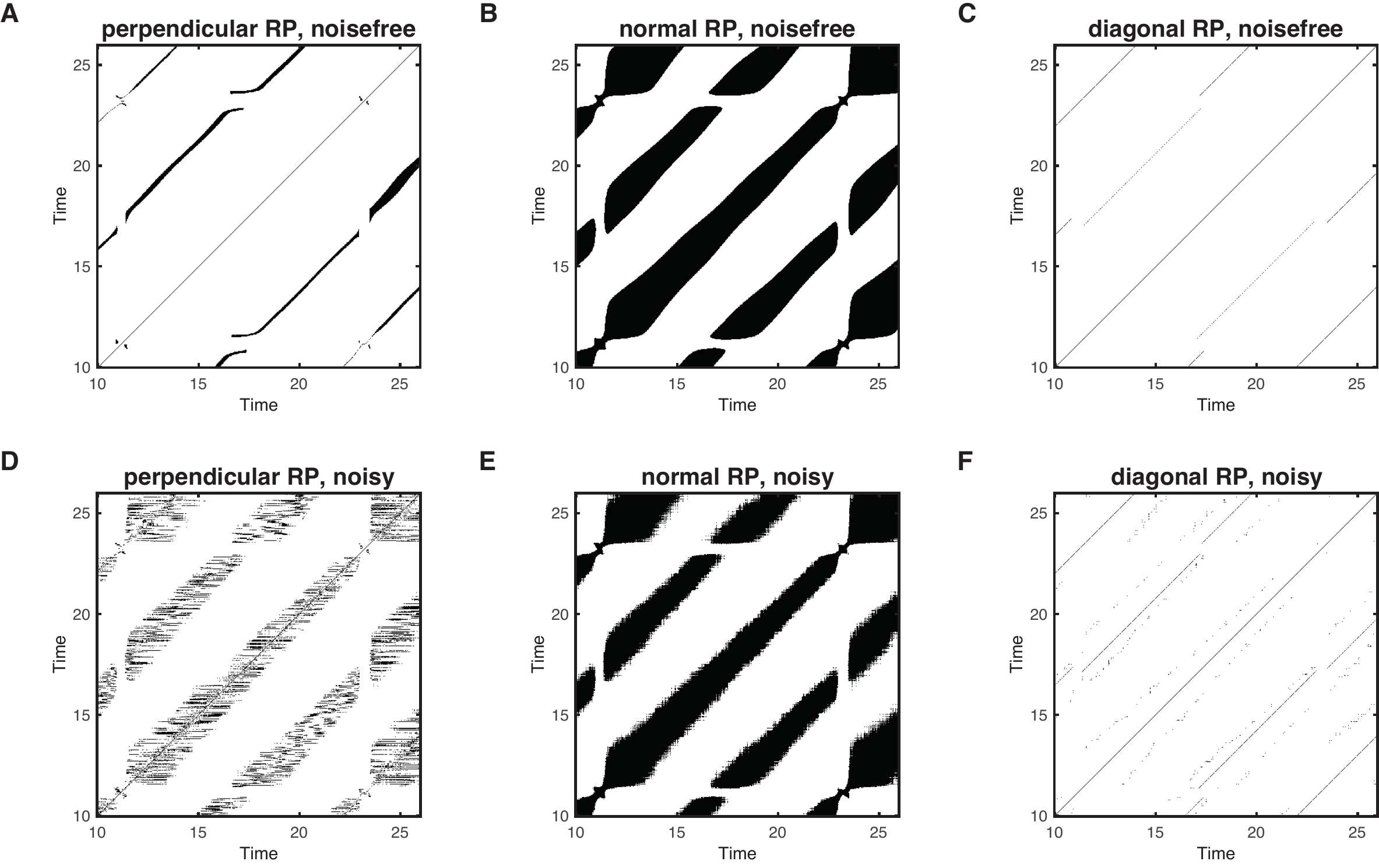}

\caption{Cut outs of (A, D) the perpendicular recurrence plot ${R}^{\perp}$, 
(B, E) normal RP, and
(C, F) the diagonal recurrence plot ${R}^{\nearrow}$ of the highly sampled 
R\"ossler system in chaotic regime (here with a sampling time of $\Delta t=0.02$). Top panels (A-C) show noise free cases, bottom panels (D-F) show their noise contaminated counterparts. Shown are results of
additive white noise as 10\% of the mean standard deviation of the multivariate signal gained from the numerical integration.
Computations have been carried out by using a fixed recurrence threshold corresponding to 35\% recurrence rate and an angle threshold
$\varphi=15^\circ$ for ${R}^{\perp}$.}\label{fig_high_sampled_noise_rps}
\end{center}
\end{figure}

For the sake of completeness and in order to investigate the behavior of our proposed methods under more 
realistic conditions, we now look at the noise corrupted R\"ossler system in the two dynamical regimes 
(Sect.~\ref{sec_results}), but with an increased sampling frequency
(sampling time $\Delta t = 0.08$) and with total lengths of the three numerically integrated time series of 
10,000 (transients already removed). 
In this setup the perpendicular recurrence plot ${R}^{\perp}$ (Sect.~\ref{sec_perp_rp}) yields meaningful results
(Fig.~\ref{fig_high_sampled_noise_rps}), and we compare its 
utility with respect to the estimation of the diagonal line length entropy to the normal RP and 
the novel diagonal recurrence plot ${R}^{\nearrow}$ (Sect.~\ref{sec_diagonal_RP}).


\begin{figure}
\begin{center}
\includegraphics[width=.85\textwidth]{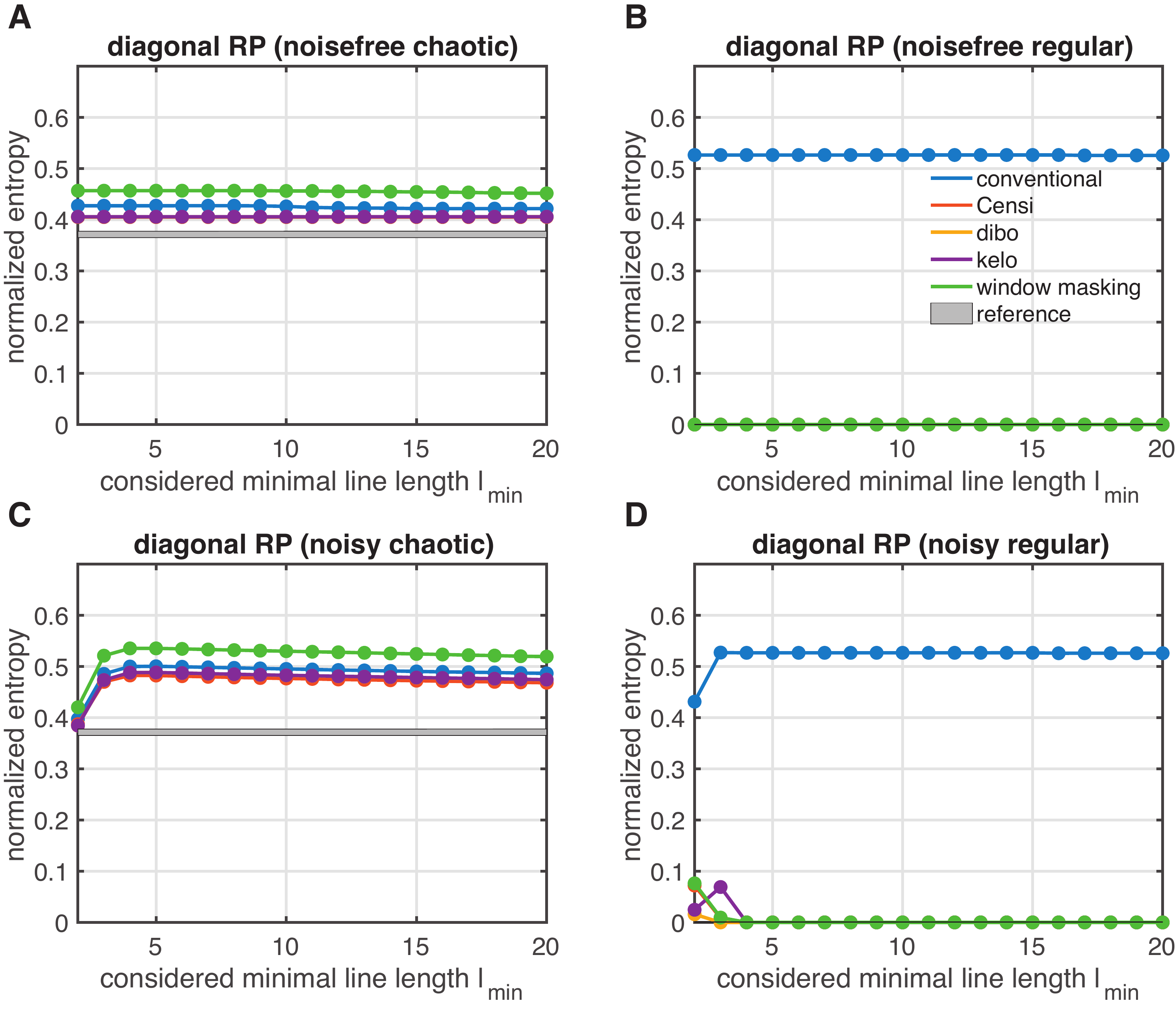}

\caption{
Normalized diagonal line length entropy estimates for all described correction schemes for counting diagonal lines 
(Sect.~\ref{sec_correction_schemes}) based on the diagonal recurrence plot ${R}^{\nearrow}$ 
(Sect.~\ref{sec_diagonal_RP}) of the high sampled R\"ossler system as a function of the 
chosen minimal line length $\ell_\text{min}$. The top panels (A - chaotic motion
, B - regular motion) show the noisefree case and in the bottom panels (C - chaotic motion, D - regular motion) 
the results for noise corrupted data are shown. We
added noise from an auto-regressive (AR) process of second order as 20\% of the
mean standard deviation of the multivariate signal gained from the numerical integration (cf. 
Eq.~\eqref{eq_ar_process}).
The underlying RPs for obtaining
${R}^{\nearrow}$ were computed using a fixed recurrence threshold corresponding to 35\% recurrence rate.
The grey shaded areas show medians of ensembles of 1,000 analytically computed reference values 
for \textit{$K_1$} $\pm$ two standard deviations of these distributions transformed by using 
Eq.~\eqref{eq_entropy_theo}.
}\label{fig_high_sampled_noise_entropy}
\end{center}
\end{figure}

Figure \ref{fig_high_sampled_noise_entropy} illustrates the capability of ${R}^{\nearrow}$
to cope with tangential motion, especially under noise. Due to a too high computational effort we
did not compute an ensemble in this case as we did in the lower sampled cases, so the 
errorbars are missing. Here we added an auto regressive (AR) process of second
order with an amplitude corresponding to 20\% of 
the mean standard deviation of the multivariate signal. 

\begin{equation}
x_{t} =a_1 x_{t-1} + a_2 x_{t-2} + \varepsilon_t \quad,
\label{eq_ar_process}
\end{equation}

with parameters $a_1=0.7$, $a_2=0.15$ and $\varepsilon_t$ denotes a white noise process with zero mean 
and constant variance of unity. Outcomes for the normal RP and the perpendicular recurrence plot ${R}^{\perp}$ can be
found in the Appendix (Figs.~\ref{fig_appendix_results_5}, \ref{fig_appendix_results_6}). 
Additive white noise of the same magnitude gave similar results to the
ones discussed here.

As expected from the examples in the last section, the diagonal RP approach
performs well under noise free conditions and all, but the conventional line counting algorithms yield zero-value entropy estimates for the
regular regime (panel B) and clearly non-zero entropies in case of the chaotic regime (panel A) close to the underestimated reference values.
The perpendicular RP also performs well in noise free conditions (Fig.~\ref{fig_appendix_results_5}). Even the conventional line length counting
leads to the desired zero entropy estimates in case of regular motion. In the presence of noise, however, ${R}^{\perp}$ is not able to distinguish
regular from chaotic behavior (Fig.~\ref{fig_appendix_results_6}), whereas ${R}^{\nearrow}$ still performs well, giving almost the same results as 
in the noise free setup. The explanation can be found in
considering the RPs (Fig.~\ref{fig_high_sampled_noise_rps}). 
For this noiselevel our proposed skeletonization approach (Fig.~\ref{fig_high_sampled_noise_rps}F) leaves small lines of maximum length 4 after its
application to the noisy normal recurrence plot (Fig.~\ref{fig_high_sampled_noise_rps}E) as noise-leftovers. The appearance of these
lines is not a result of the dynamics itself. Noise enriches the RP and its corresponding diagonal line histogram 
with small line lengths depending on the noiselevel (\cite{thiel2002}, Fig.~\ref{fig_high_sampled_noise_rps}, Fig.~\ref{fig_example_rp}A, 
Fig.~\ref{fig_example_hist_conventional}A). By increasing the minimum line length one gradually discards the
majority of the lines contained in the histogram and, thus, increases the prominence of larger line lengths
for the computation of the entropy. For a regular regime, the distribution of lines of intermediate length is
broader for all the correction schemes, but the diagonal RP. Therefore an increasing minimal line length increases
the entropy in the presence of noise for all the correction schemes, but the diagonal RP (cf. Fig.~\ref{fig_appendix_results_6}). 
In case of a chaotic regime the distribution of diagonals due to the dynamics is broader anyway (Fig.~\ref{fig_example_rp}C, 
Fig.~\ref{fig_example_hist_conventional}C) leading to the same effect. 

When increasing the minimal line length for the diagonal
RP, the entropy estimates stay more or less constant after a certain, sufficiently high, minimum line length, which depends on the
noiselevel (Fig.~\ref{fig_high_sampled_noise_entropy}C, D). The offset to the underestimated reference value for the chaotic case 
grows for increasing noiselevels. Note that the effect of additive noise is harder to tackle for the tangential motion correction schemes for
high sampled data like in this case, than it is for lower sampled examples as discussed in Sect.~\ref{sec_results}.
The higher the sampling, the finer the ramification of distance ranges $\mathfrak{D}_i$ (thickened diagonal lines).
Results for all correction schemes for a wide range of the recurrence thresholds and under the influence of white 
noise for the lower sampled situation can be
found in the Appendix (Figs.~\ref{fig_appendix_results_3} and \ref{fig_appendix_results_4}).

\section{Discussion}

In this letter we investigated the effect of the finite size of a recurrence plot on its diagonal
line length based quantification. Specifically, we showed how these border effects influence the
diagonal line length entropy and proposed three new line length counting correction schemes, which
take these effects into account (cf.~Subsects.~\ref{sec_dl2_correction},
\ref{sec_dl3_correction}, \ref{sec_windowshapes_correction}) and systematically compared them to an
already proposed correction by Censi et al. \cite{censi2004} (Subsect.~\ref{sec_censi_correction}). 
It turned out that for noise free or slightly noise corrupted map data all these
correction methods solve the problem of the biased diagonal line length entropy due to lines cut by the 
borders of the RP. However, for flow data the effect of tangential motion has a much bigger influence
on the entropy bias than the border effects. Therefore, we systematically compared already proposed
ideas to handle tangential motion and proposed a new, parameter free method, the \textit{diagonal RP}
(cf.~Sect.~\ref{sec_diagonal_RP}). 
It can
properly tackle the tangential motion effects and yield, in combination with the border effect
correction schemes, meaningful estimates for the diagonal line length entropy. We have to emphasize
that this method, in contrast to other suggested ideas, also works for noise contaminated data, is
not sensitive to the particular choice of the recurrence threshold, does not introduce any additional
parameter, and is, therefore, easy to use. In case of a noise corrupted flow-like signal the diagonal line length
entropy approaches its constant expectation value for sufficiently high choices of the minimal line length, when 
the diagonal RP together with Censi's or our propsed border effect correction schemes is used. Fairly high
recurrence thresholds (>10\% recurrence rate) favour the \textit{diagonal RP} method for intermediate or high
noise levels. 

\section*{Acknowledgments}
We thank Johannes Donath for his contribution through his Bachelor's thesis, regarding the window masking correction
scheme. This work has been financially supported by the German Research Foundation (DFG projects MA4759/8 and MA4759/9) and the European Union's Horizon 2020 Research and Innovation Programme under the Marie Sk{\l}odowska-Curie grant agreement 691037 (project QUEST).

An implementation of all the discussed correction routines (border effects and tangential motion) is available
as MATLAB code in the Zenodo archive \cite{kraemer_zenodo_2019}.


\bibliographystyle{elsarticle-num}

\bibliography{rp}

\clearpage
\appendix

\section{Sensitivity of the results to the recurrence threshold}\label{sec_appendix_sensitivity_threshold}

\begin{figure}
\begin{center}
\includegraphics[width=0.8\textwidth]{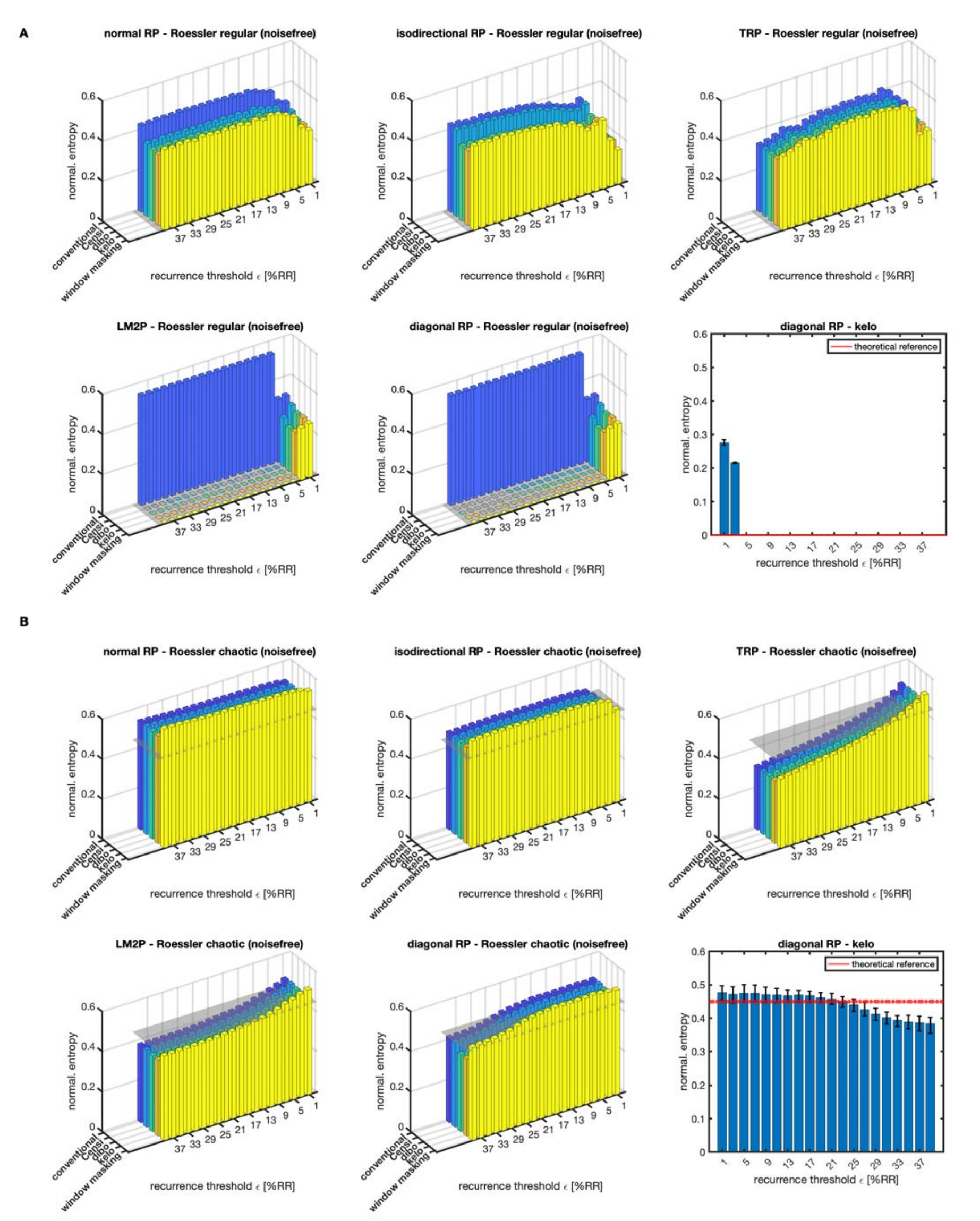}

\caption{Diagonal line length entropy estimates as a function of the recurrence threshold $\varepsilon$. Shown are results for
all described correction schemes for counting diagonal lines (Sect. \ref{sec_correction_schemes}) and suppressing tangential 
motion (Sect.
\ref{sec_tangential_motion}), except the perpendicular recurrence plot ${R}^{\perp}$. In the top panel (A) median diagonal 
line length entropy values
gained from 100 realizations of the noise free regular limit cycle regime of the R\"ossler system are shown, whereas the bottom panel
(B) shows its chaotic regime counterpart, see text in Sect.~\ref{sec_results} for details. The grey-shaded surface denotes the theoretical
expectation value (median) computed from Eq.~\eqref{eq_entropy_theo}. Results for the diagonal RP and the
kelo correction scheme are shown in the bottom right subplot, which is a cutout of the orange bars in the bottom center subplot, here
including errorbars as two standard deviations from the
computed ensemble.}\label{fig_appendix_results_1}
\end{center}
\end{figure}

\begin{figure}
\begin{center}
\includegraphics[width=0.8\textwidth]{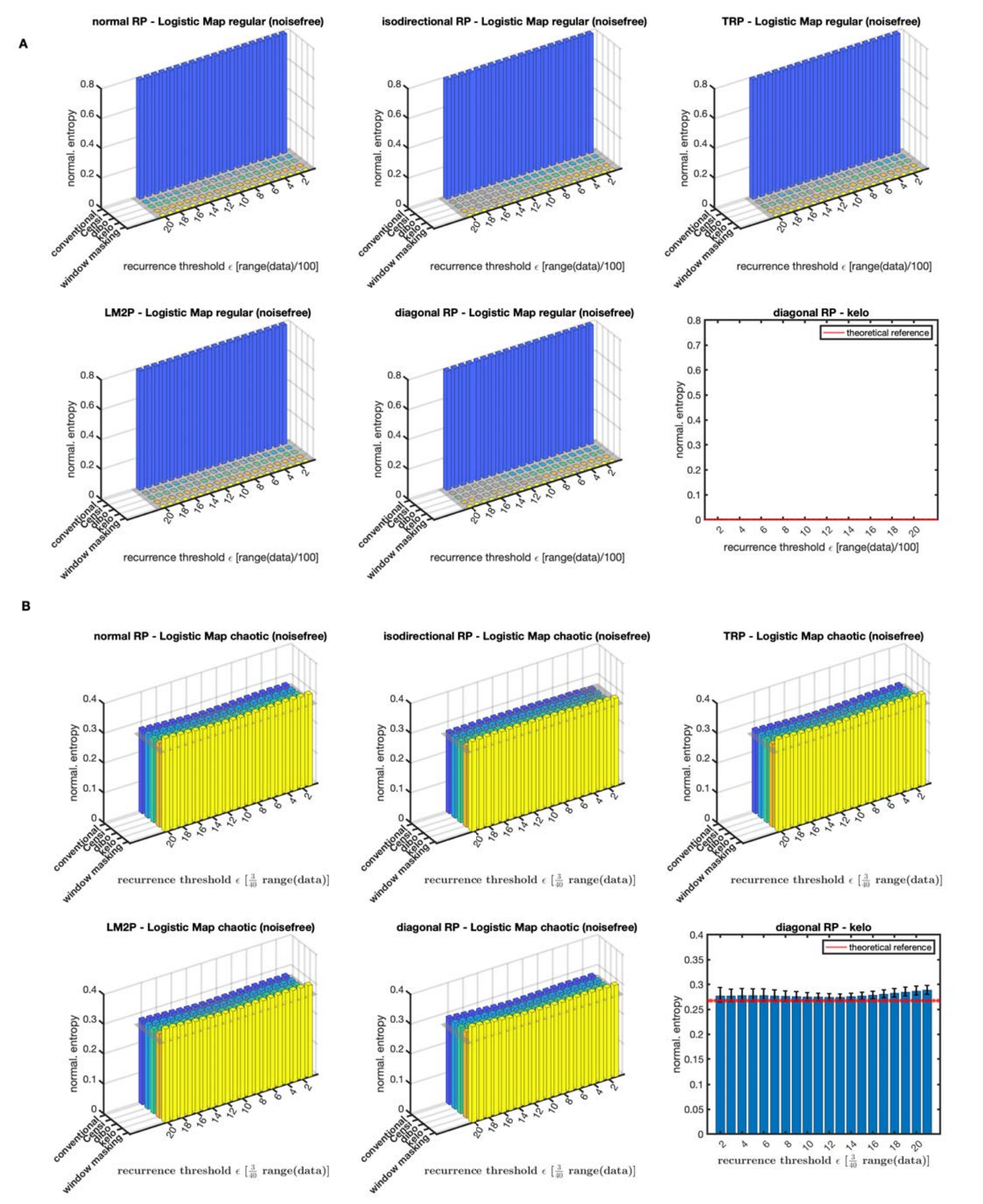}

\caption{Diagonal line length entropy estimates as a function of the recurrence threshold $\varepsilon$. Shown are results for
all described correction schemes for counting diagonal lines (Sect. \ref{sec_correction_schemes}) and suppressing tangential 
motion (Sect.~\ref{sec_tangential_motion}), except the perpendicular recurrence plot  ${R}^{\perp}$. In the top panel (A)
median diagonal line length entropy values
gained from 1,000 realizations of the noise free regular limit cycle regime of the Logistic map are shown, whereas the bottom panel
(B) shows its chaotic regime counterpart, see text in Sect.~\ref{sec_results} for details. The grey-shaded surface denotes the theoretical
expectation value (median) computed from Eq.~\eqref{eq_entropy_theo}. Results for the diagonal RP and the
kelo correction scheme are shown in the bottom right subplot, which is a cutout of the orange bars in the bottom center subplot, here
including errorbars as two standard deviations from the
computed ensemble.}\label{fig_appendix_results_2}
\end{center}
\end{figure}

\begin{figure}
\begin{center}
\includegraphics[width=0.8\textwidth]{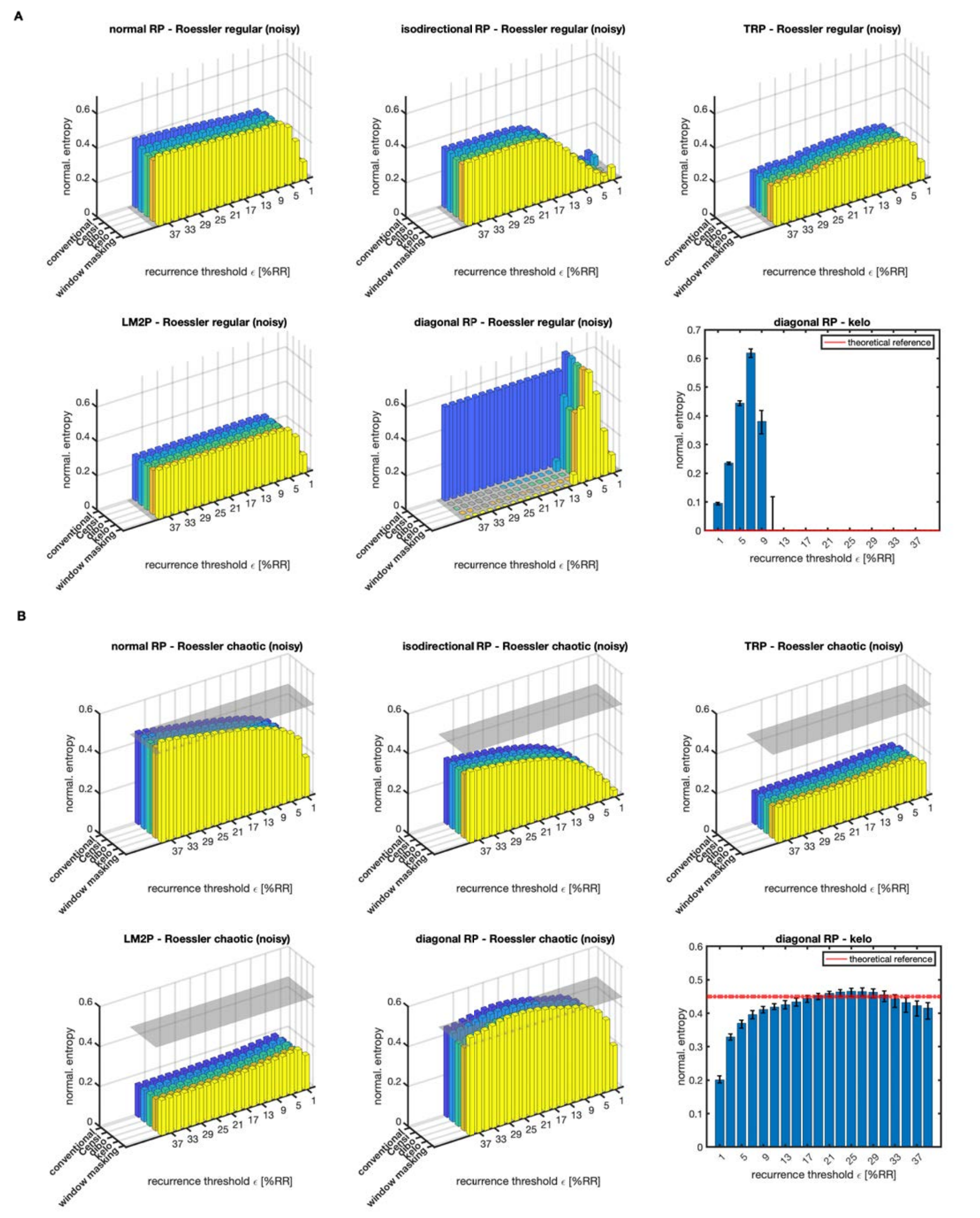}

\caption{Diagonal line length entropy estimates as a function of the recurrence threshold $\varepsilon$. Shown are results for
all described correction schemes for counting diagonal lines (Sect. \ref{sec_correction_schemes}) and suppressing tangential 
motion (Sect.~\ref{sec_tangential_motion}), except the perpendicular recurrence plot ${R}^{\perp}$. In the top panel (A) median
diagonal line
length entropy values gained from 100 realizations of the additive noise contaminated regular limit cycle regime of the R\"ossler
system are shown, whereas 
the bottom panel (B)
shows its chaotic regime counterpart, see text in Sect.~\ref{sec_results_noise} for details. Here, we added white noise as 10\% of the
mean standard deviation of the multivariate signal gained from the numerical integration. The grey-shaded surface denotes the theoretical
expectation value (median) computed from Eq.~\eqref{eq_entropy_theo}. Results for the diagonal RP and the
kelo correction scheme are shown in the bottom right subplot, which is a cutout of the orange bars in the bottom center subplot, here
including errorbars as two standard deviations from the
computed ensemble.}\label{fig_appendix_results_3}
\end{center}
\end{figure}

\begin{figure}
\begin{center}
\includegraphics[width=.8\textwidth]{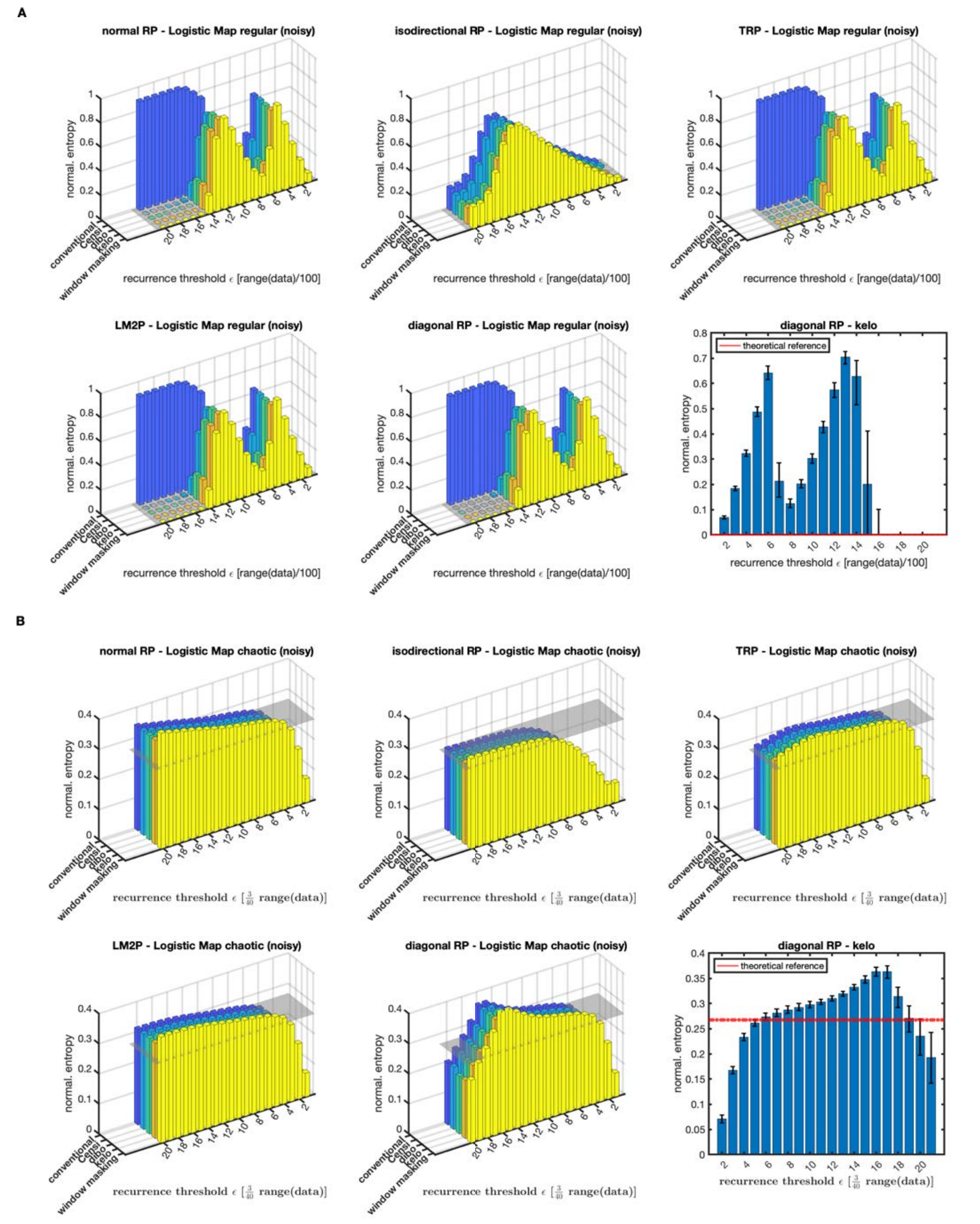}

\caption{Diagonal line length entropy estimates as a function of the recurrence threshold $\varepsilon$. Shown are results for
all described correction schemes for counting diagonal lines (Sect. \ref{sec_correction_schemes}) and suppressing tangential 
motion (Sect.~\ref{sec_tangential_motion}), except the perpendicular recurrence plot ${R}^{\perp}$. In the top panel (A) median
diagonal line length entropy values
gained from 1,000 realizations of the additive noise contaminated regular limit cycle regime of the Logistic map are shown, 
whereas the bottom panel (B)
shows its chaotic regime counterpart, see text in Sect.~\ref{sec_results_noise} for details. Here, we added white noise as 10\% of the
standard deviation of the time series. The grey-shaded surface denotes the theoretical
expectation value (median) computed from Eq.~\eqref{eq_entropy_theo}. Results for the diagonal RP and the
kelo correction scheme are shown in the bottom right subplot, which is a cutout of the orange bars in the bottom center subplot, here
including errorbars as two standard deviations from the
computed ensemble.}\label{fig_appendix_results_4}
\end{center}
\end{figure}

\clearpage
\section{Sensitivity of the results to noise}\label{sec_appendix_sensitivity_noise}


\begin{figure}
\begin{center}
\includegraphics[width=.9\textwidth]{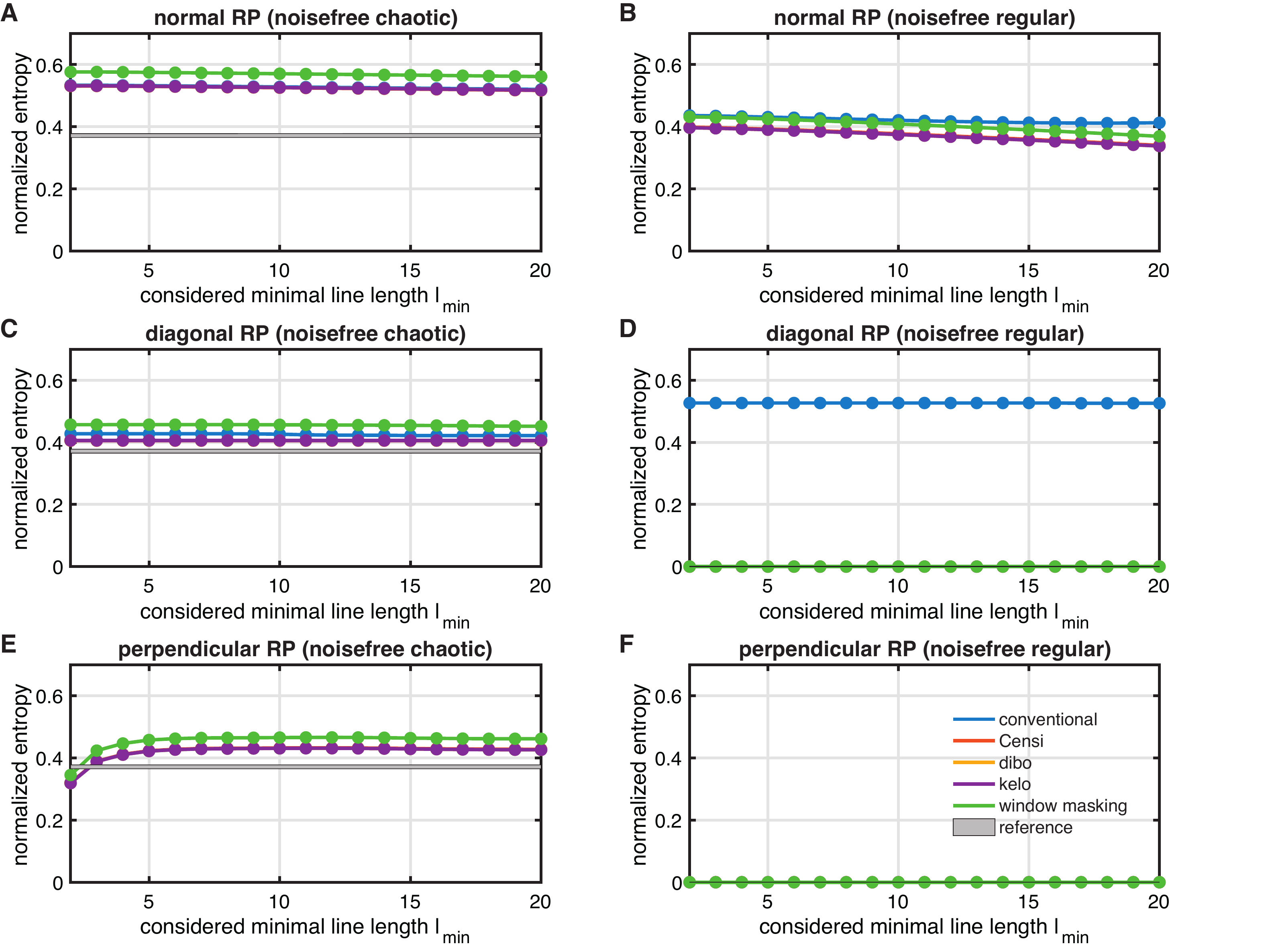}

\caption{Normalized diagonal line length entropy estimates as a function of the minimum line length $\ell_{min}$ for noisefree data from the high sampled
R\"ossler system (cf. Sect.~\ref{sec_results_noise}). In the left panels (A, C, E) the underlying system exhibits chaotic dynamics, 
whereas the right panels (B, D, F) show their regular counterparts. The normal RPs (A, B) and the perpendicular RPs (E, F) were
constructed using a fixed recurrence threshold corresponding to 35\% recurrence rate. The normal RPs served as input for obtaining 
the diagonal RPs ${R}^{\nearrow}$ (C, D) and for the computation of the perpendicular RPs ${R}^{\perp}$ we used an angle threshold
$\varphi = 15$\textdegree. The grey shaded areas show medians of ensembles of 1,000 analytically computed reference values 
for \textit{$K_1$} $\pm$ two standard deviations of these distributions transformed by using Eq.~\eqref{eq_entropy_theo}.
}\label{fig_appendix_results_5}
\end{center}
\end{figure}


\begin{figure}
\begin{center}
\includegraphics[width=.9\textwidth]{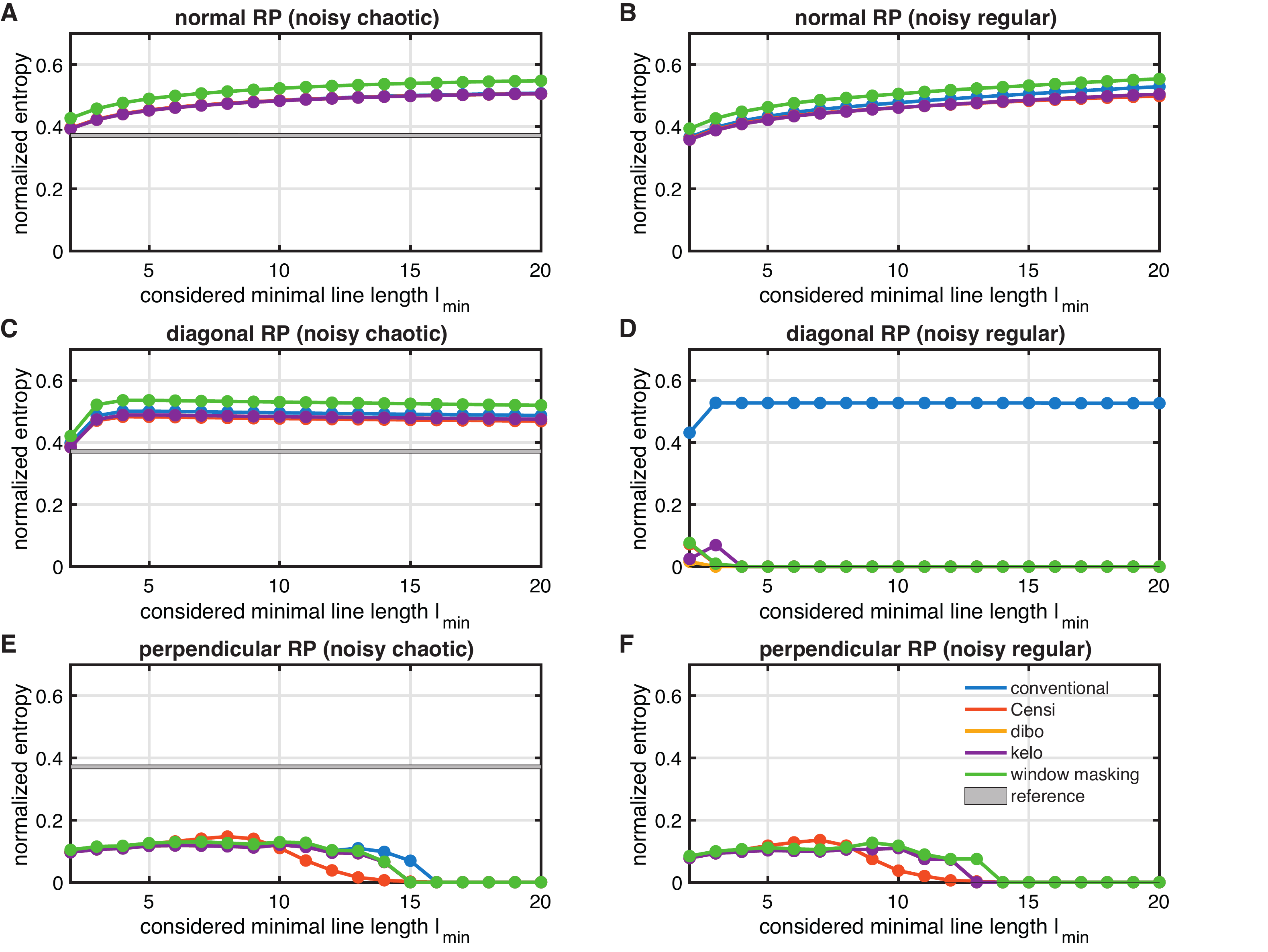}

\caption{Normalized diagonal line length entropy estimates as a function of the minimum line length $\ell_{min}$ for noise corrupted data from the high sampled
R\"ossler system (cf. Sect.~\ref{sec_results_noise}). We added noise from an auto-regressive (AR) process of second order as 20\% of the
mean standard deviation of the multivariate signal gained from the numerical integration (cf. 
Eq.~\eqref{eq_ar_process}). In the left panels (A, C, E) the underlying system exhibits chaotic dynamics, 
whereas the right panels (B, D, F) show their regular counterparts. The normal RPs (A, B) and the perpendicular RPs (E, F) were
constructed using a fixed recurrence threshold corresponding to 35\% recurrence rate. The normal RPs served as input for obtaining 
the diagonal RPs ${R}^{\nearrow}$ (C, D) and for the computation of the perpendicular RPs ${R}^{\perp}$ we used an angle threshold
$\varphi = 15$\textdegree. The grey shaded areas show medians of ensembles of 1,000 analytically computed reference values 
for \textit{$K_1$} $\pm$ two standard deviations of these distributions transformed by using 
Eq.~\eqref{eq_entropy_theo}.}\label{fig_appendix_results_6}
\end{center}
\end{figure}


\end{document}